\documentclass[review,  11pt,times]{elsarticle}
\usepackage[colorlinks=true,linkcolor=magenta, citecolor=magenta,urlcolor=magenta,anchorcolor=magenta]{hyperref}

\usepackage{pifont}
\usepackage{xcolor}

\usepackage{natbib}
\usepackage{endnotes}
\usepackage{amssymb,amsmath,amsfonts}
\usepackage{textcomp}
\usepackage{graphicx,color}
\usepackage[utf8x]{inputenc}
\usepackage{placeins}
\usepackage{graphicx}
\usepackage{caption}
\usepackage{makeidx}

\usepackage{geometry}
\usepackage{setspace}
\usepackage{amsmath,amssymb}
\usepackage{mathrsfs}
\usepackage{quoting}
\usepackage{bm}
\usepackage[normalem]{ulem}
\usepackage{amsthm}
\usepackage{subfig}

\usepackage[toc,page]{appendix}

\newcommand{\abs}[1]{\lvert#1\rvert}%

\begin{document}

\title{Morphology and flow patterns in highly asymmetric active emulsions}

\author[rvt]{G. Negro\corref{cor1}}
%{Dipartimento di Fisica, Università degli studi di Bari and INFN, Sezione di Bari, Via Amendola 173, 70126 Bari, Italy}
\ead{giuseppe.negro@ba.infn.it}

\author[rvt]{L. N. Carenza}
%{Dipartimento di Fisica, Università degli studi di Bari and INFN, Sezione di Bari, Via Amendola 173, 70126 Bari, Italy}
\ead{l.carenza@studenti.uniba.it}

\author[rvt]{P. Digregorio}
%{Dipartimento di Fisica, Università degli studi di Bari and INFN, Sezione di Bari, Via Amendola 173, 70126 Bari, Italy}
\ead{pasquale.digregorio@ba.infn.it}

\author[rvt]{G. Gonnella}
\ead{g.gonnella@ba.infn.it}
%{Dipartimento di Fisica, Università degli studi di Bari and INFN, Sezione di Bari, Via Amendola 173, 70126 Bari, Italy}
%\maketitle

\author[els]{A. Lamura}
%{Istituto Applicazioni Calcolo, CNR, Via Amendola 122/D, 70126 Bari, Italy}
\ead{a.lamura@ba.iac.cnr.it}

\cortext[cor1]{ Principal corresponding author}

\address[rvt]{Dipartimento di Fisica, Università degli studi di Bari and INFN, Sezione di Bari, Via Amendola 173, 70126 Bari, Italy}
\address[els]{Istituto Applicazioni Calcolo, CNR, Via Amendola 122/D, 70126 Bari, Italy}

\begin{abstract}
We investigate numerically, by a hybrid lattice Boltzmann method, the  morphology and the dynamics of an emulsion
made of a polar active gel, contractile or extensile, and an isotropic passive fluid. 
%The former stages  a fluid of active force dipoles, which exert either contractile or extensile stresses on their surroundings. 
We focus on the case of a highly off-symmetric ratio between the active and passive components. %with a $10:90$ ratio between the active and passive components.
In absence of any activity we observe an hexatic-ordered droplets phase, with some defects in the layout.  We study how the morphology of the system is affected by activity both in the contractile and extensile case.
In the extensile case a small amount of activity favors the elimination of defects in the array of droplets, while at higher activities, first aster-like rotating droplets appear, and then a disordered pattern occurs. In the contractile case, at sufficiently high values of activity, elongated structures are formed.
Energy and enstrophy behavior mark the transitions between the different regimes.
%Small amount of activity preserves the hexatic order while, for greater values, the hexatic pattern is progressively lost with substantial morphological differences depending on whether the emulsion is contractile or extensile.    
\end{abstract}

\maketitle
\section{Introduction}
The capability of different systems of using energy taken 
from their environment to go out of thermal equilibrium, 
gives rise to a wealth of behaviors \cite{rama2010}. 
They range from swarming, self-assembly, spontaneous
flows to other collective properties \cite{marc2013,elge2015,bech2016,sum_gonn_mar}. 
This boosted a deep interest in addressing their study in 
order to look for possible new physics, explore common features 
between different systems, and develop new strategies 
in designing synthetic devices and materials with smart properties.

Self-propelled objects represent a remarkable example of
active matter. Starting from the seminal model of Vicsek \cite{vics1995} 
for swarms, it was later realized that common features can be
traced in several systems at different scales promoting the
introduction of statistical models able to describe such behaviors
\cite{doi:10.1146/annurev-conmatphys-031214-014710,PhysRevE.90.052130,PhysRevLett.108.248101,PhysRevLett.110.055701}. Another example of active matter, sharing many
properties with suspensions of swimmers, is made by active gels
that have been introduced to model mixtures of polar biological
filaments with (active) motor proteins \cite{prost2015active,joanny2009active,2ecce8df3f7d44d789984888c34bef7a}. Their continuum modeling is based on
the liquid-cristal description of long filaments
in the nematic phase supplemented with additional contributions
to introduce the motor activity \cite{tone2005,rama2010,marc2013}.
% \cite{dege1995}

Research in this field has been mainly focused on single-component active systems and to a lesser extent on the behavior of solutions of active and passive components.
Mixtures of self propelled and passive particles have been studied by Brownian-like simulations \cite{mccandlish2012spontaneous,PhysRevE.92.032118}, focusing on the role of activity in separating the two components of the mixtures.
Binary fluids with an active component have been considered in \cite{tjhu2012,blow2014biphasic} showing that the active part may cause instabilities on active passive interface. Recently \cite{bone2017} a model has been introduced where emulsification of the active component is favored by the presence of surfactant added to the mixture. This model generalizes the aforementioned active gel theory to describe the behavior of a mixture of isotropic passive and polar active fluids.  The goal was to have a system with a tunable amount of active material that can be dispersed homogeneously in  the fluid. This would also represent a further important step in the study of active turbulent fluids \cite{thampi2016active,PhysRevX.5.031003} (with the possibility of tuning the intensity but also the spatial distribution of energy input in the system).

%Different routes can be thought to
%produce systems with such features. Embedding active particles 
%in (passive) water-in-oil emulsions \cite{sanc2012} offers a first chance while
%another possibility is given by water solutions of self-attractive
%cytoskeletal gels or sticky bacteria, where attractions can be tuned
%by depletion forces \cite{schw2012} and the arrest of phase separation
%can be triggered by either a colloidal surfactant or hydrodynamic
%\cite{scag2016} and steric stabilization \cite{schw2012}.

In \cite{bone2017}  a symmetric mixture of active and passive components was considered. Their equilibrium configurations are dominated by the formation of local ordered lamellae \cite{doi:10.1063/1.3077004,Corberi,Aiguo,xu2006morphologies,doi:10.1063/1.3143183}. It was shown that that activity may modify such a configuration 
leading to  a variety of morphologies 
whose formation strongly depends on the intensity and the kind of active doping. 
Indeed, polar active fluids are said to be either extensile (\emph{e.g.} bacterial colonies and microtubules bundles) or contractile (\emph{e.g.} actine and myosin filaments) according to the nature of the stress exerted by the active component on its neighborhood. Furthermore, intensity of active doping can be tuned by keeping under control the amount of fuel available to active particles. \textcolor{black}{
This corresponds experimentally to keeping under control the amount of ATP in  active gels of bundled microtubles \cite{Henkin20140142} or the amount of oxygen available, the concentration of ingredients, or the temperature in bacterial suspensions. In the present model this is done by
introducing a parameter representing the strength of the active stress acting in the system (see Section 2).}
The main result was that, even if under symmetric conditions, activity modifies lamellar configurations into an emulsion of  passive droplets in an active matrix at sufficiently high contractile activity. On the other hand, lamellae change their morphology into rotating active droplets in the extensile case.

In this work we complement the previous analysis by considering a highly off-symmetric mixture with a $10:90$ ratio of the active and passive components  both for extensile and contractile systems. 
Here the equilibrium state of the fluid is characterized by an ordered array of droplets of the minority phase positioned at the vertices of a triangular lattice. We will show  that, 
despite the strong unbalance between the two components, activity greatly affects the morphology of the system, leading to the development of a wide range of patterns both for the concentration and the velocity field.
In the extensile case a small amount of activity favors the elimination of defects in the system, as shown by measuring the number of defects in Voronoi tessellation. By increasing activity, isolated droplets tend to merge forming
larger 'islands' of active material and then, at still larger activity, 
 big rotating droplets are observed. In the contractile case  activity
promotes the rupture of the hexagonal phase and the appearance of a matrix of 
the active component in the passive flowing background, differently from what happened in the symmetric case. The morphological study is supported by the analysis of the kinetic energy and enstrophy behavior.   

The dynamic equations for the concentration of the active material and the 
polarization, which fixes the average orientation of the active component, are derived from a 
proper free-energy functional and supplemented with the Navier-Stokes 
equations for the whole fluid in the incompressible limit.
These equations are numerically solved by using a hybrid lattice Boltzmann
method \cite{tjhu2012,tiri2009,tiri2011,cate2009,dema2014,tjhu2015}.

The paper is organized in the following way. The next Section is devoted to
present the thermo-hydrodynamic description of the system, the numerical method,
and the set of parameters relevant for the present study.
In Sections 3 and 4 the results for the morphology and the corresponding flow
patterns will be shown and related to the observed behaviors of 
energy and enstrophy. Finally, a discussion with some remarks and possible 
future lines of investigation will conclude the paper.

\section{Model}

We outline here the hydrodynamic model and the numerical method used to conduct our study. We consider a fluid comprising a mixture of active material and solvent
with total mass density $\rho$.
The physics of the resulting composite material can be described by using an extended version of the well-established active gel theory~\cite{rama2010,marc2013,tjhu2012,tjhu2015,elsen,joannyPRL}.
%The order is described via a vector $\bf P$, called the polarisation field.
%which ensures that there is no head-tail simmetry, as the in the case of standard passive nematic liquid crystals.  
The hydrodynamic variables are the density of the fluid $\rho$, its velocity ${\bf v}$, the  concentration of the active material $\phi$,
 and the polarization ${\bf P}$, which determines the average orientation of the active material.
The dynamic equations ruling the evolution of the system are
\begin{eqnarray}
%\nabla\cdot\mathbf{v} & = & 0,\label{cont}\\
\rho\left(\frac{\partial}{\partial t}+\mathbf{v}\cdot\nabla\right)\mathbf{v} & = & -\nabla p+ \nabla\cdot\underline{\underline{\sigma}}^{total}\ ,\label{nav}\\
\frac{\partial \phi}{\partial t}+\nabla\cdot\left(\phi\mathbf{v}\right)&=&\nabla \cdot \left( M\nabla\frac{\delta F}{\delta \phi}\right),\label{conc_eq}\\
\frac{\partial\mathbf{P}}{\partial t}+\left(\mathbf{v}\cdot\nabla\right)\mathbf{P}&=&-\underline{\underline{\Omega}}\cdot\mathbf{P}+\xi\underline{\underline{D}}\cdot\mathbf{P}%\nonumber\\
-\frac{1}{\Gamma}\frac{\delta F}{\delta\mathbf{P}},\label{P_eq}
\end{eqnarray}
in the limit of incompressible fluid. The first one is the Navier-Stokes equation, where $p$ is the isotropic pressure and $\underline{\underline{\sigma}}^{total}$ is the total stress tensor \cite{BerisEdwards}. 
Eqs.~(\ref{conc_eq})-(\ref{P_eq}) govern the time evolution of the concentration of the active material and of the polarization field, respectively.
Since the amount of active material is locally conserved, the time evolution of the concentration field can be written  as a convection-diffusion equation, Eq.~(\ref{conc_eq}), where $M$ is the mobility, $F$ a free  energy functional as defined later, 
and $\delta F/\delta\phi$ is the chemical potential.
%The former is a convection-diffusion equation in which $M$ is the mobility and $\mu=\delta F/\delta\phi$ is the chemical potential. 
The dynamics of the polarization field follows an advection-relaxation equation, Eq.~(\ref{P_eq}), borrowed from polar liquid crystal theory. Here $\Gamma$ is the rotational viscosity, {$\xi$ is a constant controlling the aspect ratio of active particles (positive for rod-like particles and negative for disk-like ones)},
$h=\delta F/\delta\textbf{P}$ is the molecular field. $\underline{\underline{D}}=(\underline{\underline{W}}+\underline{\underline{W}}^T)/2$ and  
$\underline{\underline{\Omega}}=(\underline{\underline{W}}-\underline{\underline{W}}^T)/2$ represent the symmetric and the antisymmetric parts 
of the velocity gradient tensor $W_{\alpha\beta}=\partial_{\beta}v_{\alpha}$, where Greek indexes denote Cartesian components. These contributions are in addition to the material derivative as the liquid crystal can be rotated or aligned by the fluid~\cite{BerisEdwards}.
The stress tensor $\underline{\underline{\sigma}}^{total}$ considered in the Navier-Stokes equation of the model, Eq.~(\ref{nav}), 
is splitted into the equilibrium/passive and non-equilibrium/active part:
\begin{equation}
\underline{\underline{\sigma}}^{\textit{total}}=\underline{\underline{\sigma}}^{\textit{passive}}+\underline{\underline{\sigma}}^{\textit{active}} \mbox{.}
\end{equation}
The passive part represents elastic response from solutes and is, in turn, the
sum of three terms:
\begin{equation}
\underline{\underline{\sigma}}^{\textit{passive}}=\underline{\underline{\sigma}}^{\textit{viscous}}+\underline{\underline{\sigma}}^{\textit{elastic}}+\underline{\underline{\sigma}}^{\textit{interface}} \mbox{.}
\end{equation}
The first term is the viscous stress written as  $\sigma_{\alpha\beta}^{viscous}=\eta(\partial_{\alpha}v_{\beta}+\partial_{\beta}v_{\alpha})$ where 
$\eta$ is the shear viscosity.
The second term is the elastic stress analogous to the one used in nematic liquid crystal hydrodynamics ~\cite{BerisEdwards}:
\begin{eqnarray}
\sigma_{\alpha\beta}^{elastic}=&&\frac{1}{2}(P_{\alpha}h_{\beta}-P_{\beta}h_{\alpha})-\frac{\xi}{2}(P_{\alpha}h_{\beta}+P_{\beta}h_{\alpha})\nonumber\\
                           &&-\kappa\partial_{\alpha}P_{\gamma}\partial_{\beta}P_{\gamma}\label{eq:elastic-stress},
\end{eqnarray}
%where $\textbf{P}$ is the polarisation, $\textbf{h}=\delta F/\delta\textbf{P}$ is the molecular field~\cite{deGennes} and the parameter $\xi$ controls whether the liquid crystal 
%molecules are rod-like shaped ($\xi>0$) or disk-like shaped ($\xi<0$). 
where $\kappa$ is the elastic constant of the liquid crystal and the parameter $\xi$ depends on the geometry, as already mentioned. In addition $\xi$ establishes whether the fluid is flow aligning ($|\xi|>1$) or flow tumbling ($|\xi|<1$) under shear. 
The third term is borrowed from binary mixtures theory. It includes interfacial contribution between the passive and the active phase:   
\begin{equation}
\sigma_{\alpha\beta}^{interface}=\left( f-\phi\frac{\delta F}{\delta\phi} \right)\delta_{\alpha\beta} - \frac{\partial f}{\partial\left(\partial_{\beta}\phi\right)} \partial_{\alpha}\phi \ .
\end{equation}
Here $f$ is the free energy density. 
The active contribution to the stress tensor, not stemming from the free energy, is given by~\cite{elsen,Simha}
\begin{equation}
\sigma_{\alpha\beta}^{active}=-\zeta \phi \left(P_{\alpha}P_{\beta}-\frac{1}{3}|{\bf P}|^2\delta_{\alpha\beta}\right)\label{eq:active-stress},
\end{equation}
where $\zeta$ is the activity strength that is positive for extensile systems (pushers) and negative for contractile ones (pullers). 
The active stress drives the system out of equilibrium by injecting energy into it and satisfies the symmetry ${\bf{P}}\rightarrow - {\bf P}$.
%It should be noted that  in the derivation of the active stress tensor higher order terms (such as $\partial_{\alpha}P_{\beta}$) are allowed by symmetry. Our assumption is anyway acceptable as the degree of asymmetry of elongated fibres (e.g. actin filaments) is supposed to be small.

The thermodynamics properties of the binary mixture, in absence of activity, are encoded in the following free-energy functional that couples the Landau-Brazovskii model ~\cite{braz} to the distortion free-energy of a polar system:
\begin{eqnarray}\label{fe}
&F&[\phi,\mathbf{P}]
=\int d\mathbf{r}\,\{\frac{a}{4\phi_{cr}^4}\phi^{2}(\phi-\phi_0)^2+\frac{k}{2}\left|\nabla \phi\right|^{2}+\frac{c}{2}(\nabla^2\phi)^2 \nonumber\\
&-&\frac{\alpha}{2} \frac{(\phi-\phi_{cr})}{\phi_{cr}}\left|\mathbf{P}\right|^2+ \frac{\alpha}{4}\left|\mathbf{P}\right|^{4}+\frac{\kappa}{2}(\nabla\mathbf{P})^{2}
+\beta\mathbf{P}\cdot\nabla\phi\} \ \ .
\end{eqnarray}
%Here $\phi$ is a scalar order parameter representing the concentration of the active material and $\mathbf{P}$ is a vector field representing
%its average orientation.
This is a generalization of the free energy functional for active binary mixtures defined in \cite{elsen}.
The first term, multiplied by the phenomenological constant $a>0$, describes the bulk properties of the fluid, 
the second and third ones determine the interfacial tension. 
Notice that here a negative value of $k$ favors the formation of interfaces while a positive value of $c$ has to guarantee the stability of the free-energy \cite{braz}. The Landau-Brazovskii model, with only the $\phi$ terms in the first line of Eq. (\ref{fe}), when the composition is symmetric, has a transition line for $k_{cr}=-\sqrt{8c \tilde{a} /15}$, in the mean field approximation \cite{PhysRevA.46.4836}, where $\tilde{a}=a\phi_0^2/8\phi_{cr}^4$ is half the coefficient of the $\phi$ quadratic term in \eqref{fe}. Lowering $k$ from positive to negative values leads the system to move from pure ferromagnetic phase to configurations where interfaces between components are favored; for values lower than $k_{cr}$ the system exhibits a periodic behavior such that the equilibrium state of the system is characterized by lamellae. For asymmetric compositions  droplets of the minority phase are stable \cite{doi:10.1063/1.3077004,Gompper_critic,Or_Gon_ye}.  
The bulk term is chosen in order to create two free energy minima, one ($\phi=0$) corresponding to the passive material and the other one ($\phi\simeq\phi_0$) corresponding to the active phase; $\phi_{cr}=\phi_0/2$, where $\phi_{cr}$ is the critical concentration for the transition from isotropic ($|{\bf P}|=0$) to polar ($|{\bf P}|>0$) states.
%The third term, multiplied by the positive constant $c$, guarantees the stability of the free energy.
\begin{figure}[t!]
\centering
 \subfloat[][ \label{img:no_activity_phi}]
{\includegraphics[height=.278\textheight]{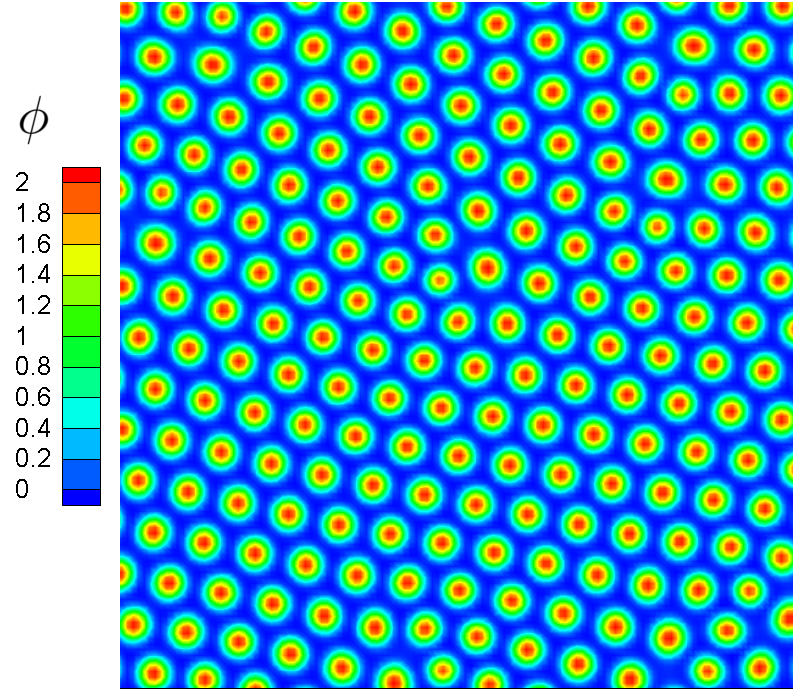}} \ \ \ 
\subfloat[][ \label{img:no_activity_vor}]
{\includegraphics[height=.278\textheight]{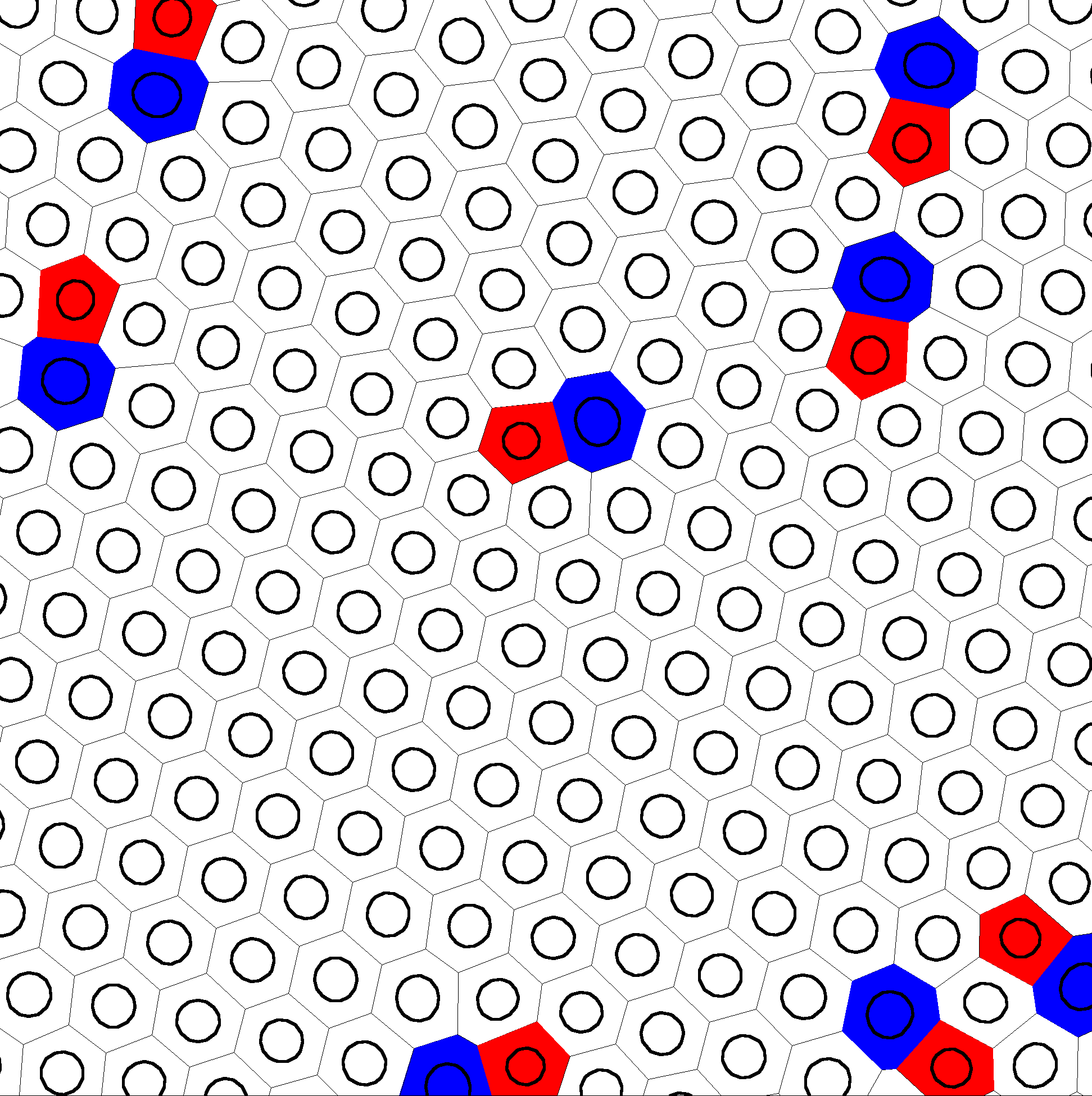}}\\
\caption{\textbf{(a)} Snapshot of $\phi$ contour plot for a configuration  in the stationary regime at $\zeta=0$, 
for a portion of size $L=128$, of a system of original size $L=256$. The color code displayed here is the same of all the contour plots in the rest of the work. \textbf{(b)} Voronoi tessellation for the same configuration in \textbf{(a)}. }
\end{figure}
The bulk properties of the polar liquid crystal are instead controlled by the $|\textbf{P}|^2$ and $|\mathbf{P}|^{4}$ terms, multiplied by the positive constant $\alpha$. The choice of $\phi_{cr}$ has been made in order to break the symmetry between the two phases and confine the polarization field in the active phase $\phi > \phi_{cr}$.
The second last term proportional to $(\nabla\mathbf{P})^{2}$ describes the energy cost due to elastic deformation in the liquid crystalline phase, gauged by the elastic constant $\kappa$ (in the single elastic constant approximation). Finally, the last term takes into account the orientation of the polarization at the interface of the fluid. If $\beta\ne 0$, $\textbf{P}$ preferentially points perpendicularly to the interface (normal anchoring): towards the passive (active) phase if $\beta>0$ ($\beta<0$). \\
\begin{figure}[t!]
\centering
\subfloat[][ \label{img:0_006_1}]
{\includegraphics[width=.227\textwidth]{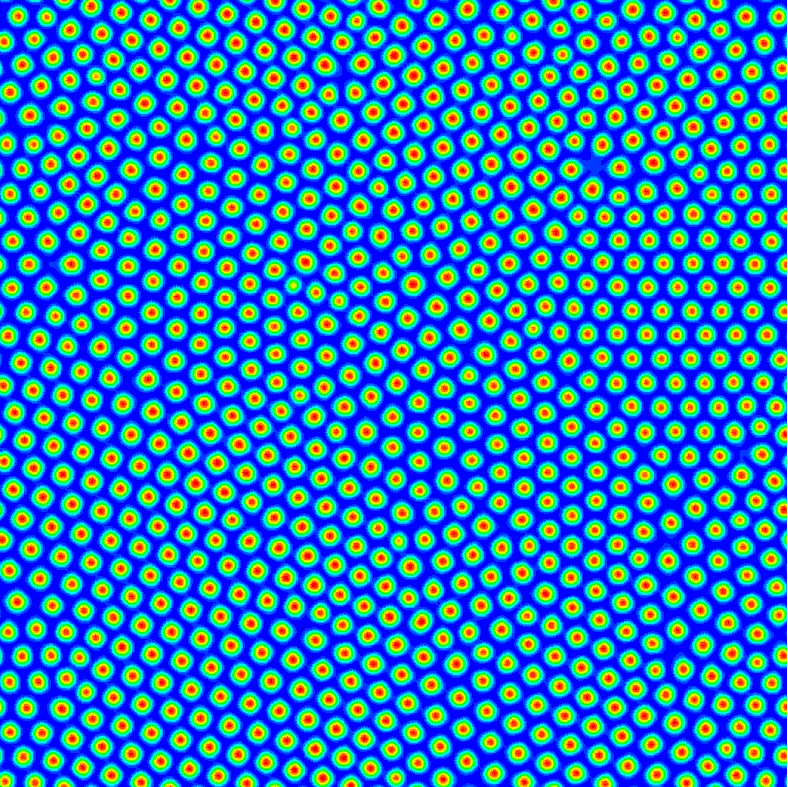}} \
\subfloat[][ \label{img:0_006_2}]
{\includegraphics[width=.227\textwidth]{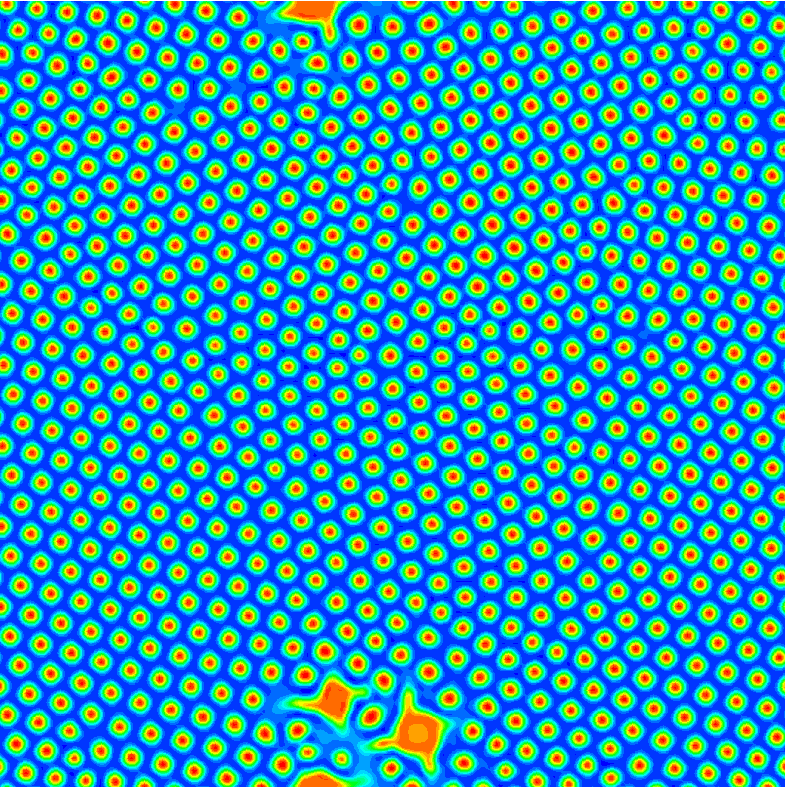}} \
\subfloat[][ \label{img:0_006_3}]
{\includegraphics[width=.23\textwidth]{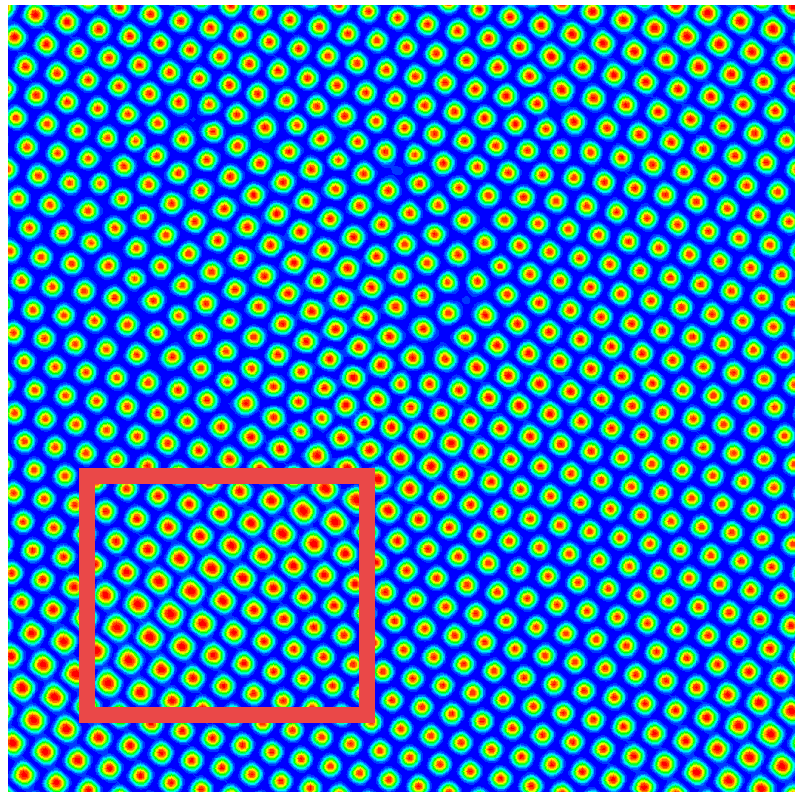}} \
\subfloat[][ \label{img:0_006_4}]
{\includegraphics[width=.226\textwidth]{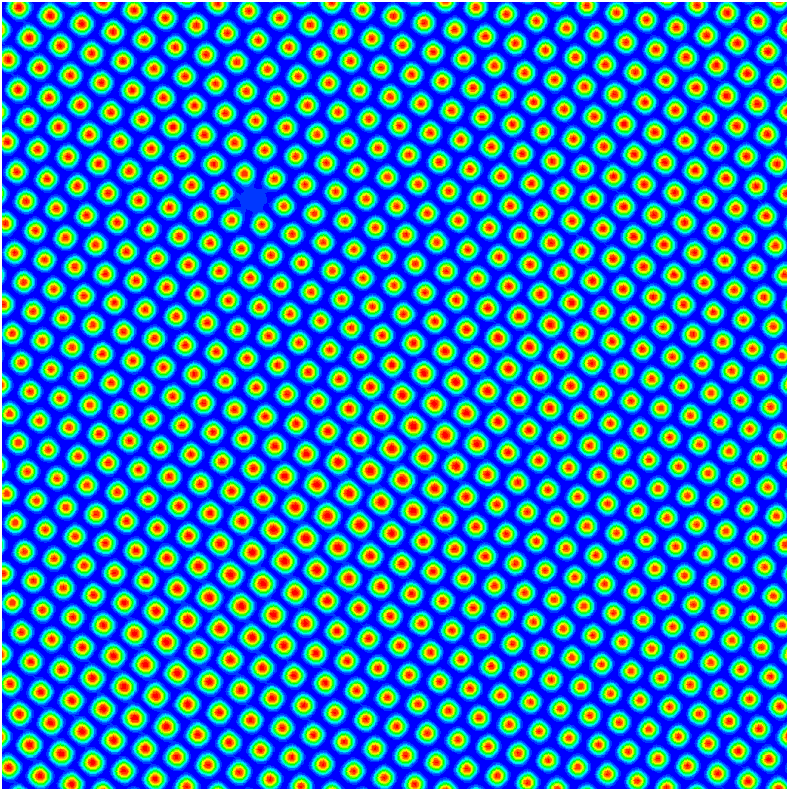}} \\
\subfloat[][ \label{img:vor_0_006_1}]
{\includegraphics[width=.226\textwidth]{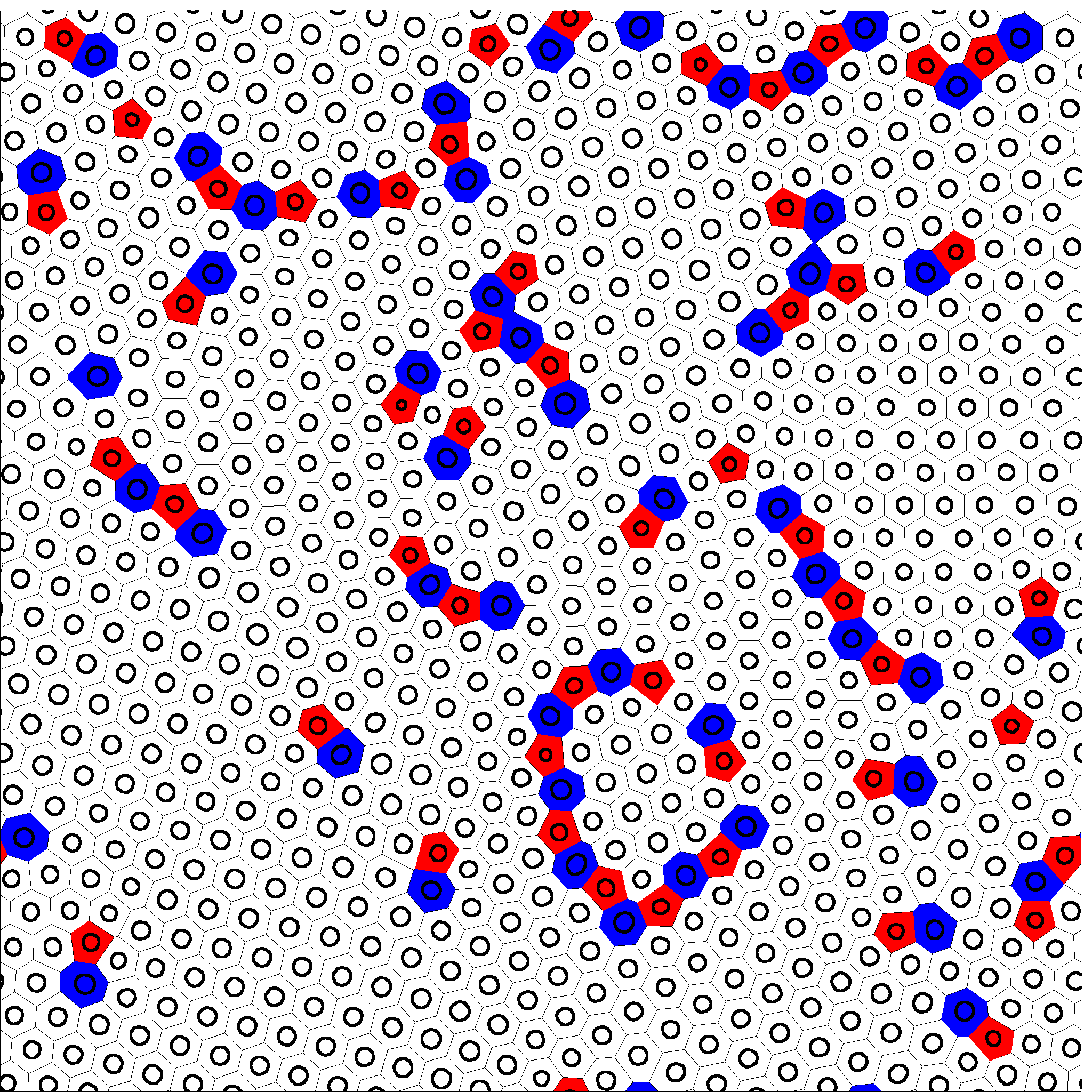}} \ \
\subfloat[][ \label{img:vor_0_006_2}]
{\includegraphics[width=.226\textwidth]{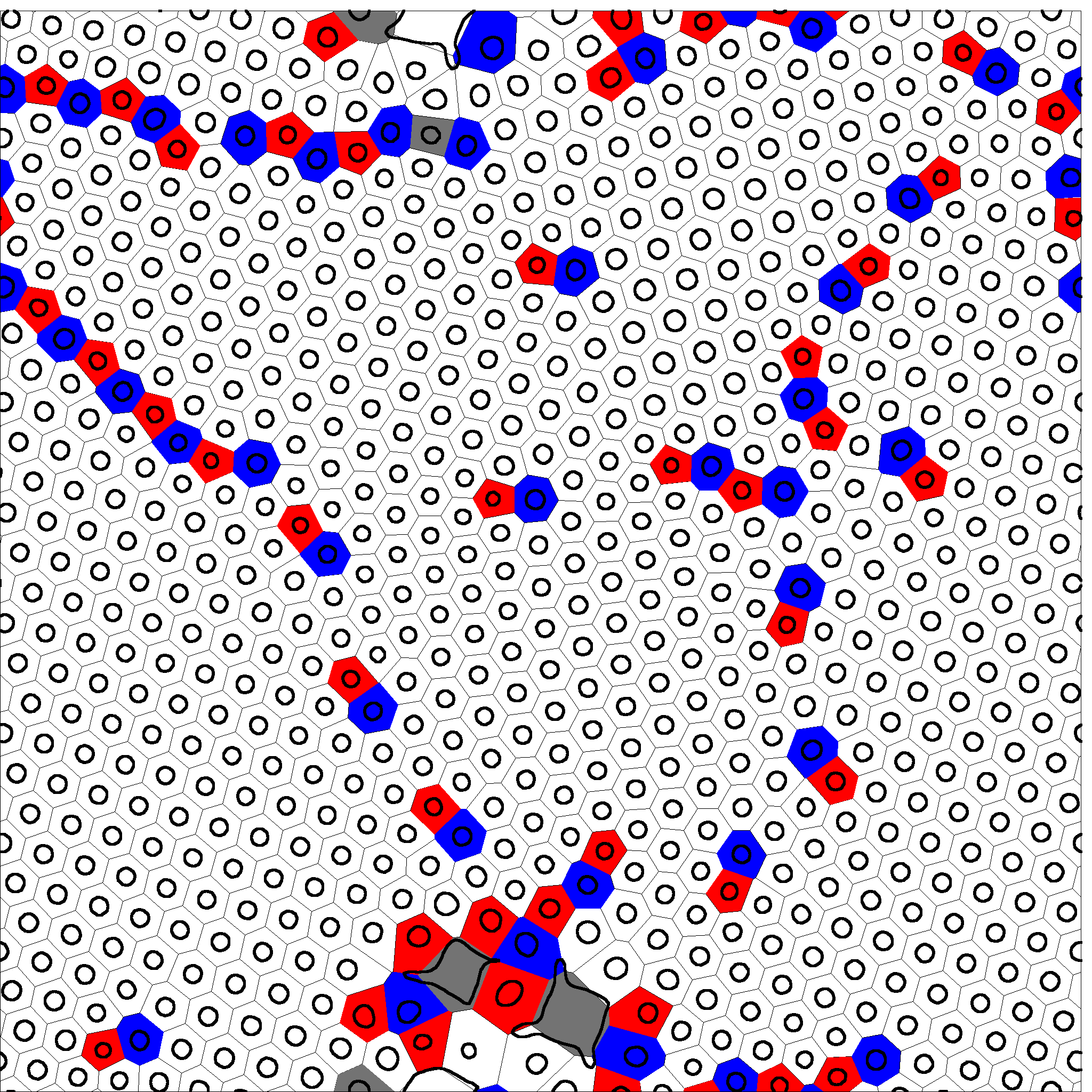}} \ \
\subfloat[][ \label{img:vor_0_006_3}]
{\includegraphics[width=.226\textwidth]{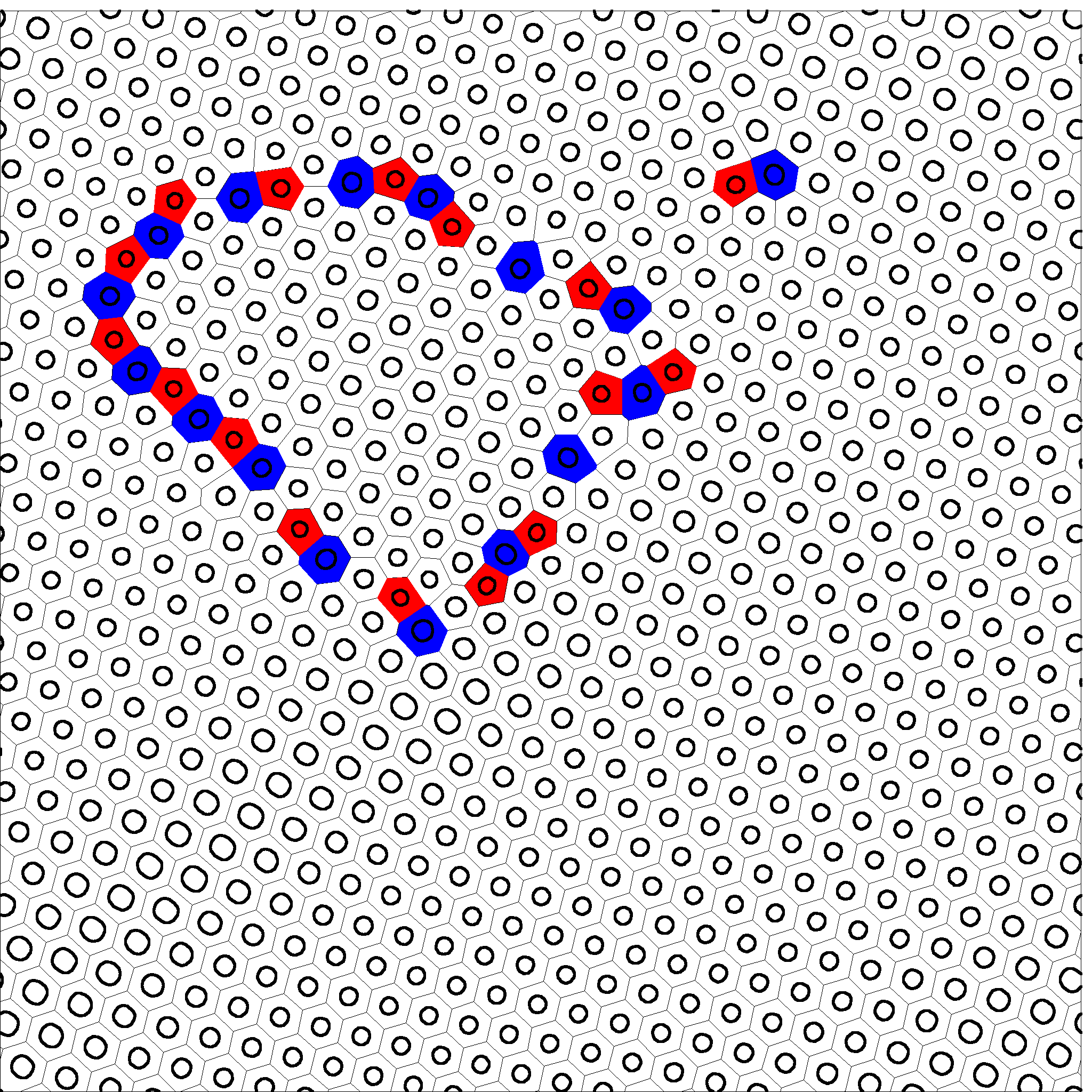}} \ \
\subfloat[][ \label{img:vor_0_006_4}]
{\includegraphics[width=.226\textwidth]{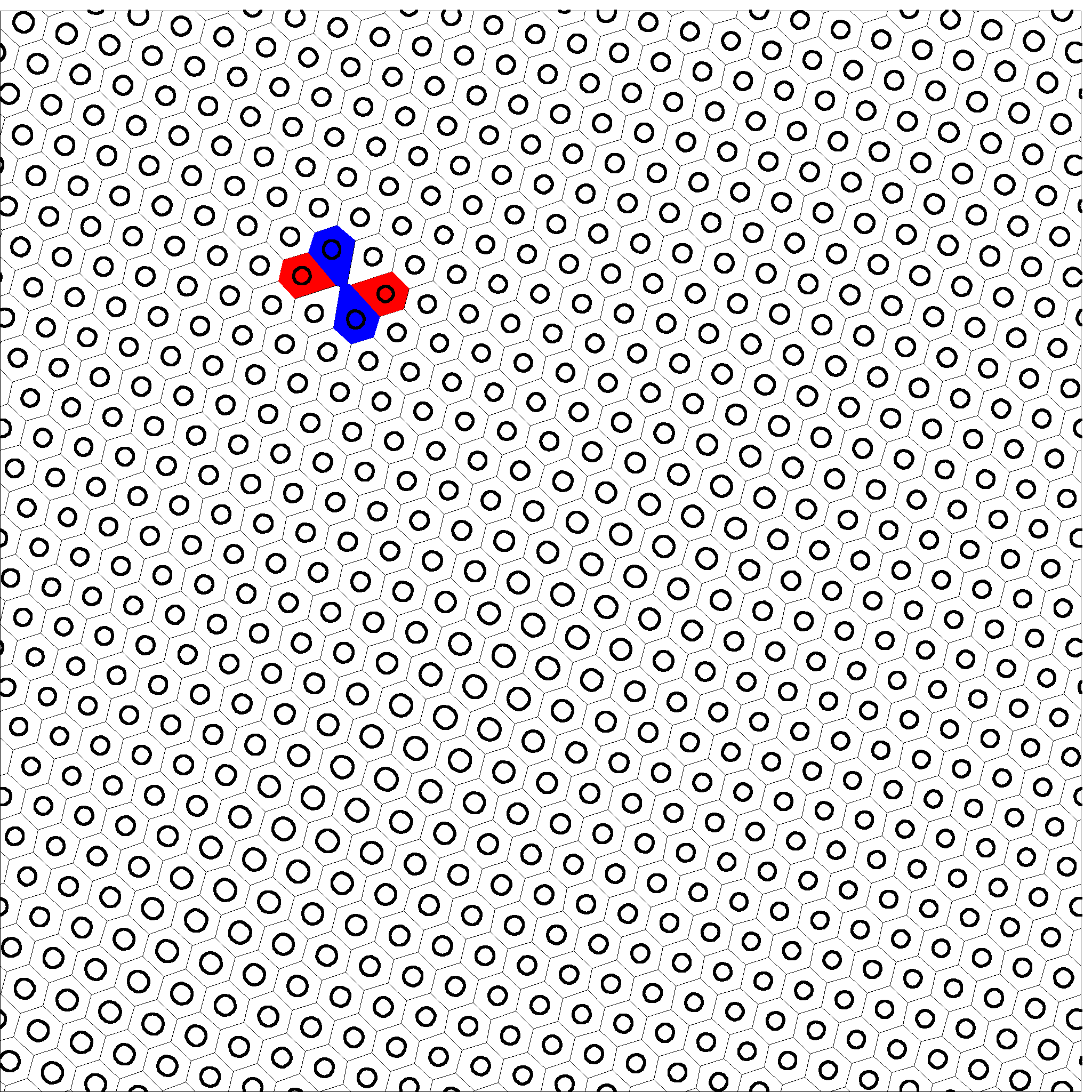}}
\caption{\textcolor{black}{Snapshots of $\phi$ contour plots (upper panels), and corresponding Voronoi tessellation (lower panels) at times $2\times10^{5}$ (a), \ $12\times10^{5}$ (b), \ $40\times10^{5}$ (c), \ $62\times10^{5}$ (d) for  $\zeta=0.006$ on a lattice of size $L=256$. The box in Fig. \ref{img:0_006_3} indicates a region in which the droplet size is bigger compared to the rest of the system. The final configuration (Fig. \ref{img:vor_0_006_4}) is almost-completely hexatically ordered as witnessed by the presence of very few defects. These latter ones are in correspondence of the hole visible in in Fig. \ref{img:0_006_4} in the same position.}}
\label{fig:contour_plot_extensile_vor_0.006}
\end{figure}
The equations of motion of the exotic polar active emulsion, Eqs.~(\ref{nav})-(\ref{P_eq}),  are solved by means of a hybrid lattice Boltzmann (LB) scheme, which combines a LB treatment for the Navier–Stokes equation (see~\ref{appendix1} for more details) with a finite-difference predictor-corrector algorithm to solve the order parameter dynamics. 
%In fact, due to the coupling between velocity concentration and polarization fields in Eqs.~(\ref{P_eq}) and~(\ref{conc_eq}), 
\begin{figure}[t!]
\centering
{\includegraphics[width=.7\textwidth]{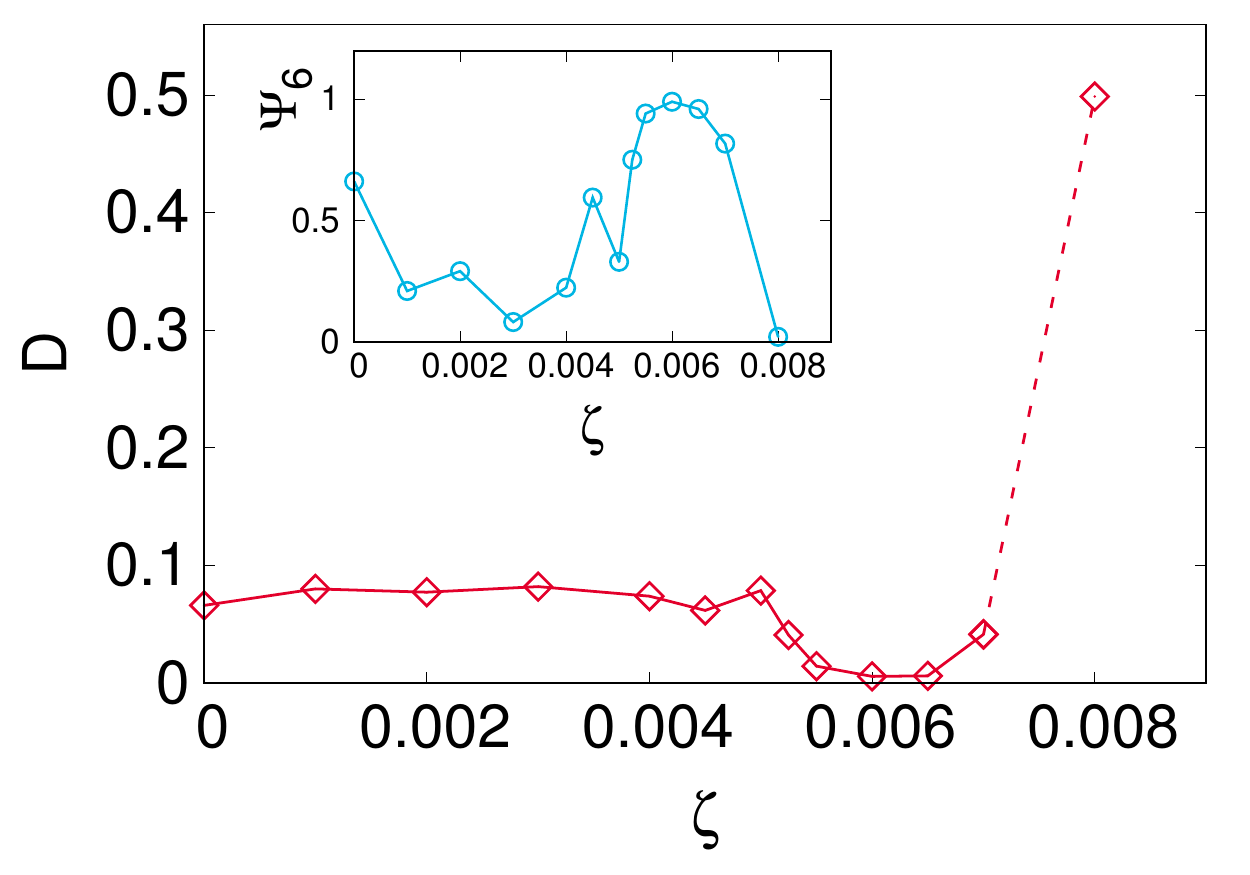}} 
\caption{\textbf{(Main figure)} Defects ratio (droplets without 6 neighbors over the total number of droplets in the configuration) vs activity, in the stationary time regime, for systems of size $L=256$. The figurative dashed line stands for the fact that increasing activity it is no more possible a coherent defects analysis, due to the formation of asters first and  completely non definite structures for strong activity (see next figures). \textbf{(Inset)} \textcolor{black}{Global hexatic order parameter (\ref{eq:hex_par}) as a function of activity.}}
\label{fig:defects_stat}
\end{figure}
%The LB scheme is coupled with a finite-difference predictor-corrector algorithm, in order to solve simultaneously Eqs~(\ref{nav})-(\ref{P_eq}).
%This ensures better convergence characteristics 
Simulations have been performed on a periodic square lattice of size $L=256$.
The concentration $\phi$ ranges from $\phi\simeq 0$ (passive phase) to 
$\phi\simeq 2$ (active phase).
Unless otherwise stated, parameter values are $a=4\times 10^{-3}$, $k=-6\times 10^{-3}$, $c=10^{-2}$, $\alpha=10^{-3}$, $\kappa=10^{-2}$, $\beta=0.01$,
$\Gamma=1$, $\xi=1.1$, $\phi_0=2.0$, and $\eta=1.67$. All quantities in the text are reported in lattice units.\\
We have considered a $10:90$ mixture of the active and passive component following two initialization procedures. In one case we started from a random configuration of polarization and concentration. The latter has been initialized considering a random variation of ten percent around its average value. Starting from this configuration, the system has been equilibrated without activity. Then, activity has been switched on and the evolution of the system studied.
In the other procedure the random initial configuration has been evolved in the presence of activity.
We checked that the two procedures lead to the same behaviour at late times.
The results presented in the following generally are obtained with the second initialization procedure.
%%%%%%%%%%%%%%%%%%%%%%%%%%%%%%%%%%%%%%%%%%%%%%%%%% FINE SEZIONE MODELLO 
\section{Extensile case}
\subsection{Small activity, hexatic order, and defects}
\begin{figure}[t!]
\centering
\subfloat[][\emph{$\zeta=0.0072$} \label{zeta=0.0072}]
{\includegraphics[width=.271\textwidth]{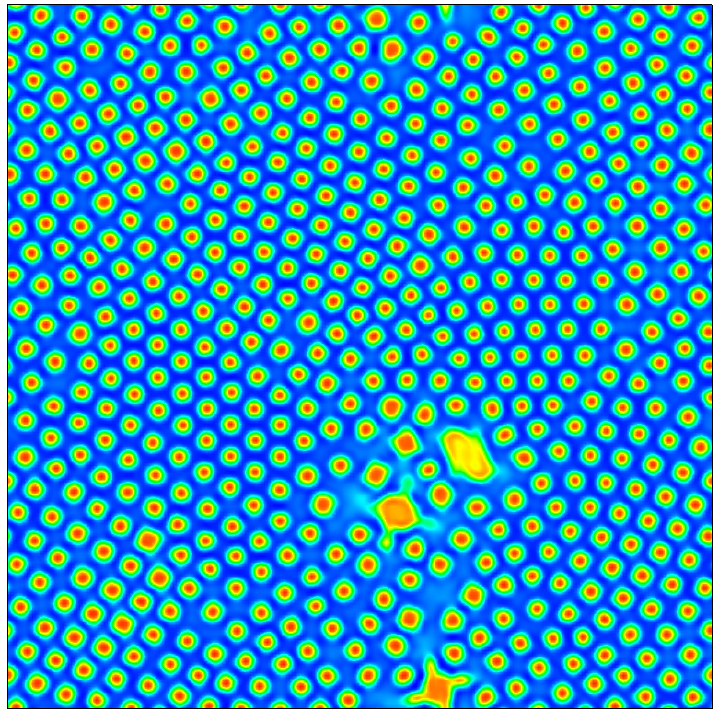}} \
\subfloat[][\emph{$\zeta=0.0074$} \label{zeta=0.0074}]
{\includegraphics[width=.271\textwidth]{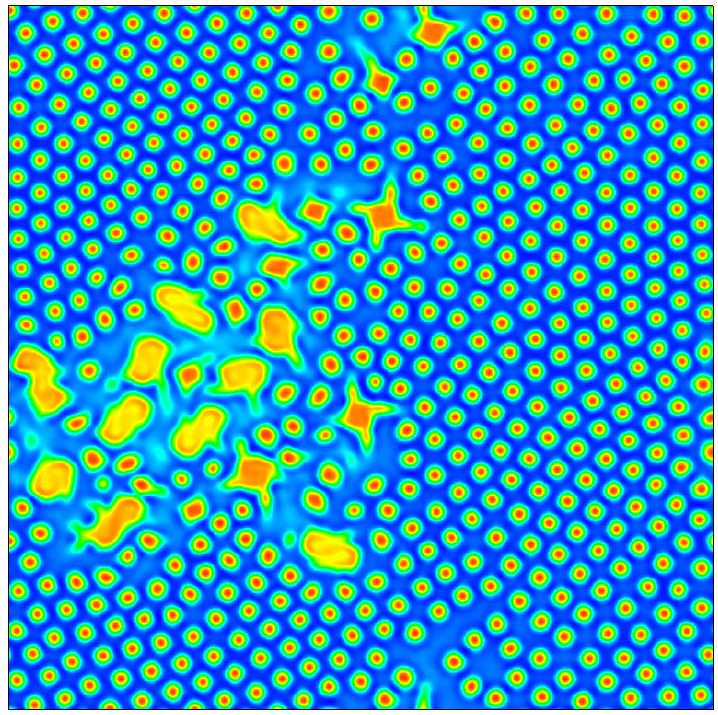}} \
\subfloat[][\emph{$\zeta=0.0076$} \label{zeta=0.0076}]
{\includegraphics[width=.271\textwidth]{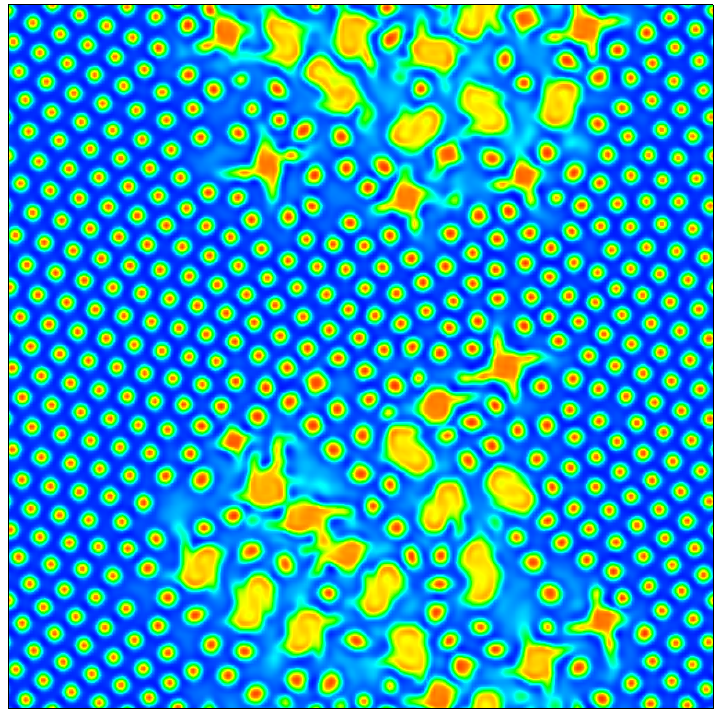}} \\
\subfloat[][\emph{$\zeta=0.0078$} \label{zeta=0.0078}]
{\includegraphics[width=.271\textwidth]{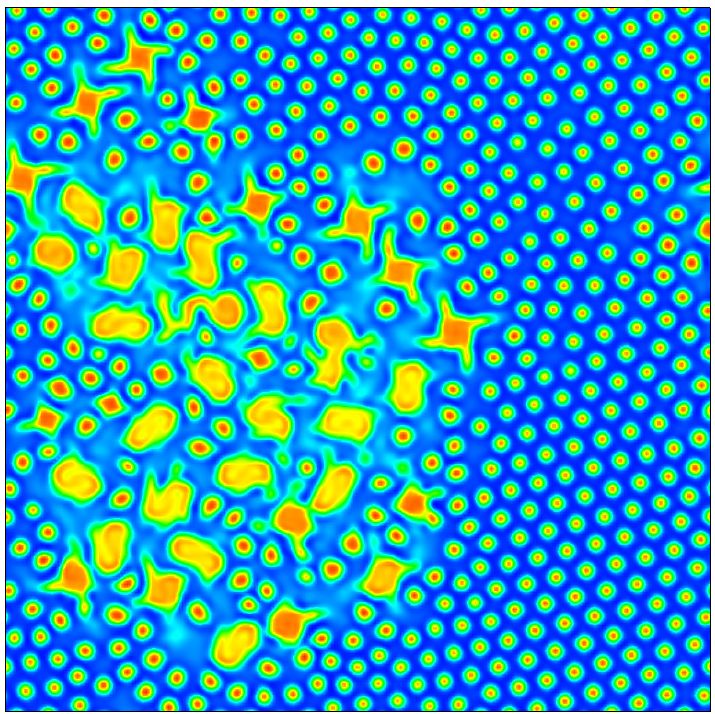}} \
\subfloat[][\emph{$\zeta=0.0080$} \label{zeta=0.0080}]
{\includegraphics[width=.271\textwidth]{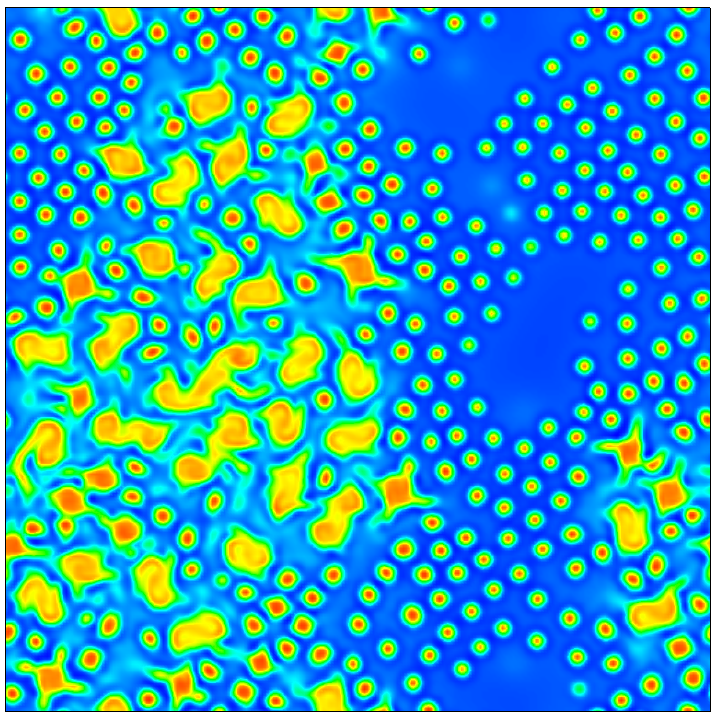}} \
\subfloat[][\emph{$\zeta=0.0090$} \label{zeta=0.0090}]
{\includegraphics[width=.271\textwidth]{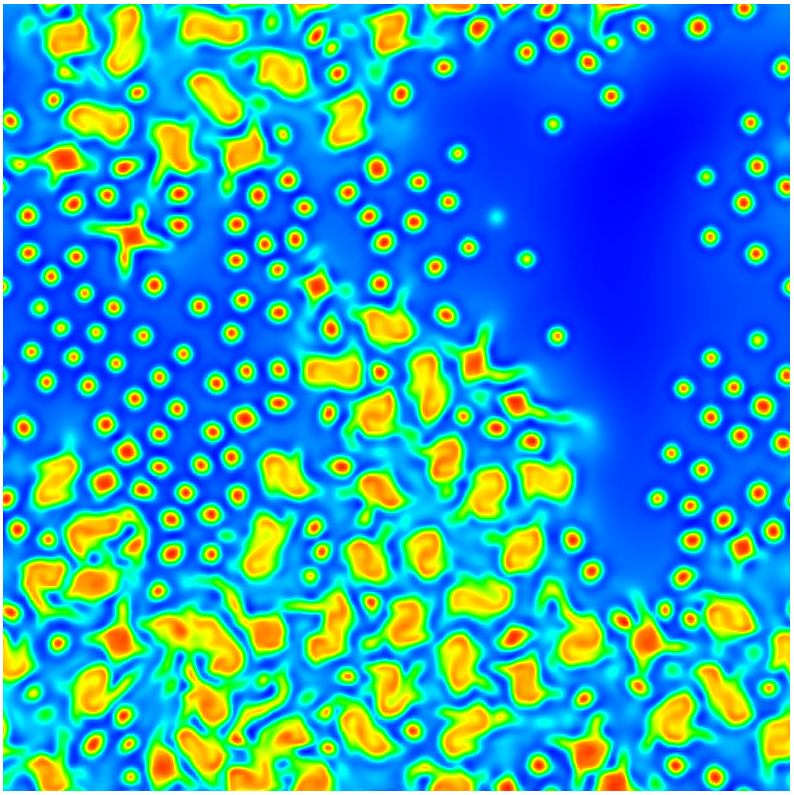}} \
\caption{Snapshots of $\phi$ contour plots at late times for different choices of $\zeta>0$, for a lattice of size $L=256$.}
\label{fig:SEQUENCE_FINO}
\end{figure}

Having in mind to fully characterize the behavior of the system as the value of activity changes, we begin by considering the case when the activity is off.
In this case, at equilibrium, the system is characterized by an ordered array of droplets as can be seen looking at the contour plot of the concentration $\phi$ in Fig.~\ref{img:no_activity_phi}. The droplets (and their centers of mass) can be easily pinpointed by putting a proper cutoff on the concentration field to distinguish active regions from passive ones. Each closed region of lattice sites that falls beyond the cutoff, is identified as a droplet. A good choice for the cutoff with our choice of the parameters is seen to be $\phi \sim 1.5$. 
Droplets are  hexatically ordered, that is they occupy vertices of a triangular lattice, besides the presence of some defects. A \textit{Voronoi tessellation} is used in order to unambiguously identify the nearest-neighbors network for the centers of mass of each droplet.
Voronoi tessellation establishes a partitioning of the space with one closed region for each center of mass, according to the following rule: the region associated to the $i$-th droplet contains all the points of the space that are closer to its center of mass than to any other droplets. In Fig.~\ref{img:no_activity_vor} it is shown the use of this analysis. Droplets with 5 nearest neighbors are highlighted in red, while those with 7 neighbors in blue. Observe that most of the defects appear in pairs indicating the presence of dislocations in the system\cite{chaikin2000principles}.

%On the right of figure \ref{img:no_activity_vor} the \textcolor{blue}{cutted-off} concentration of the active material is reported, in order to better appreciate the sahpe and the size of droplets.
%We will use these plots in this section in order to qualitatively characterize the system in the case of small activity.

%\textcolor{blue}{description of figures of droplets to better understand the shape(cut-off) voronoi tasselation for defects.}
%\textcolor{red}{The system, in the absence of activity, would arrange in an exatic ordered configuration with some defects(\ref{img:no_activity_vor}).}

%\begin{figure}[t!]
%\centering
%\subfloat[][\emph{$t=10^6$}]
%{\includegraphics[width=.4\textwidth]{Plot_Morphology/seq2.png}}\ 
%\subfloat[][\emph{$t=1.2\times 10^6$}]
%{\includegraphics[width=.402\textwidth]{Plot_Morphology/seq3.png}} \ \ 
%\caption{Formation of aster-like droplets structure at $\zeta=0.0072$, for a system of size $L=256$.  }
%\label{fig:contour_plot_evolution_0.007}
%\end{figure}
For non-zero, but still small (positive) values of $\zeta$, the \textit{hexatic-droplet} phase survives. At sufficiently high values of $\zeta$, but still in the small activity regime, activity is able to reduce the number of defects, driving the system in an almost completely ordered phase, as shown in the time evolution of Fig.~\ref{fig:contour_plot_extensile_vor_0.006}, at $\zeta=0.006$. 
The system first forms transient differently-oriented domains bounded by grain-boundary defects (an example of a closed grain boundary is shown in Fig.~\ref{img:vor_0_006_3}), then they shrink during the time evolution and disappear at the end, leaving a single hole in the layout.
\textcolor{black}{We checked taht if instead of switching on activity from the beggining, we start from configurations equilibrated without activity, and then we switch it on, the defects dynamics is not significantly affected.}
In the course of this  evolution we also observe other features, here only transient, that play a major role at larger activity.
In Fig.~\ref{img:0_006_2} few larger aster-like droplets are observed. We refer as \textit{aster-like droplets}, big rotating droplets or simply \emph{asters} to non-circular droplets which have the shape of an aster associated to the formation of vortices in the velocity field, as will be seen later. The presence of these aster-like droplets will be a predominant feature in the large activity regime as will be seen further in the text.
We observe also a slight increase of droplets size during time evolution of the system as  highlighted in Fig. \ref{img:0_006_3}.
\begin{figure}[t!]
\centering
%{\includegraphics[width=.85\textwidth]{positive_activity/energy_vs_time_0072.png}} \
{\includegraphics[width=.85\textwidth]{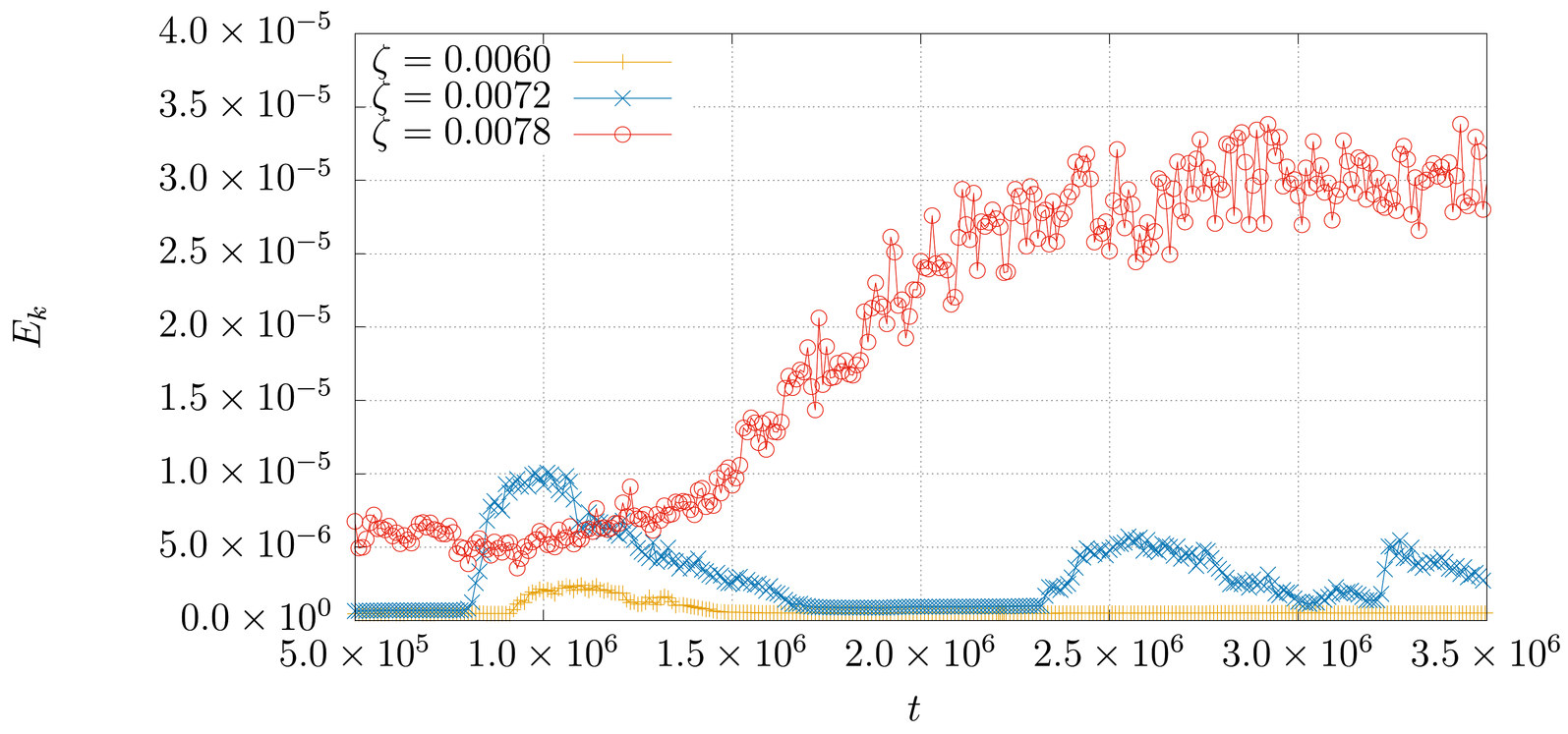}} \
\caption{Time evolution of kinetic energy for different values of $\zeta$. The \textbf{middle blue} curve is the time evolution for the same system of Fig. \ref{zeta=0.0072}
. For $\zeta=0.006$ (\textbf{bottom yellow} curve) the peak corresponds to the formation of asters as explained in the text but at late times the stationary state is characterized by an hexatic order and the kinetic energy goes to zero. The \textbf{top red} curve corresponds to the case $\zeta=0.0078$ and is displayed in order to show that the system reach a stationary state. }
\label{fig:en.evolution}
\end{figure} 
At the end, for $\zeta=0.006$ and slightly larger values of $\zeta$, activity is such to let the droplets rearrange in an almost defect-free final configuration.
To better understand the effect of activity on the hexatic arrangement we plot in Fig.~\ref{fig:defects_stat} the defects ratio $D$ (with respect to the number of droplets) as the activity varies. We see that for very small values of activity, the hexatic order survives but there is no appreciable change in the number of defects with respect to the passive limit. 
However, for values of $\zeta$ around $0.006$, or slightly larger, for the longest simulated times, the hexatic order is enhanced as activity is such to remove the defects.
At larger values of $\zeta$  a change in the overall morphological behavior of the mixture will be observed and the defects analysis looses significance.
{\color{black} In order to further characterize order in the droplet phase we use a \textit{hexatic order parameter}, as the one used for studying hexatic order and phase transitions in two-dimensional systems \cite{PhysRevLett.41.121}. Local hexatic order parameter and its global average are defined, as in \cite{PhysRevLett.119.268002}, by}
\begin{equation}
	\psi_{6i} = \frac{1}{N_i} \sum_{j=1}^{N_i} \exp{6\mbox{i}\theta_{ij}} \mbox{,} \qquad \Psi_6=\frac{1}{N} \biggl| \sum_{i=1}^{N} \psi_{6i} \biggr| \mbox{,}
	\label{eq:hex_par}
\end{equation}
{\color{black}where $N_i$ is the number of neighbors of the $i$-th droplet, $\theta_{ij}$ is the angle that droplets $i$ and $j$ form with a reference axis, and $N$ is the total number of droplets.
As shown in the inset of Fig.~\ref{fig:defects_stat}, the measure of $\Psi_6$ is consistent with the picture found by looking at the number of defects. As activity is increased from $0$, the global hexatic parameter fluctuates around a positive value, which means that the 
system is partially hexatically ordered except for some defects. 
Within the activity range where we see no defects, $\Psi_6$ reaches its saturation value 1, being the system in a perfectly ordered phase.}
\begin{figure}[t!]
\centering
\subfloat[][\emph{$\zeta=0.0030$}\label{img:zoom0.0030}]
{\includegraphics[width=.3\textwidth]{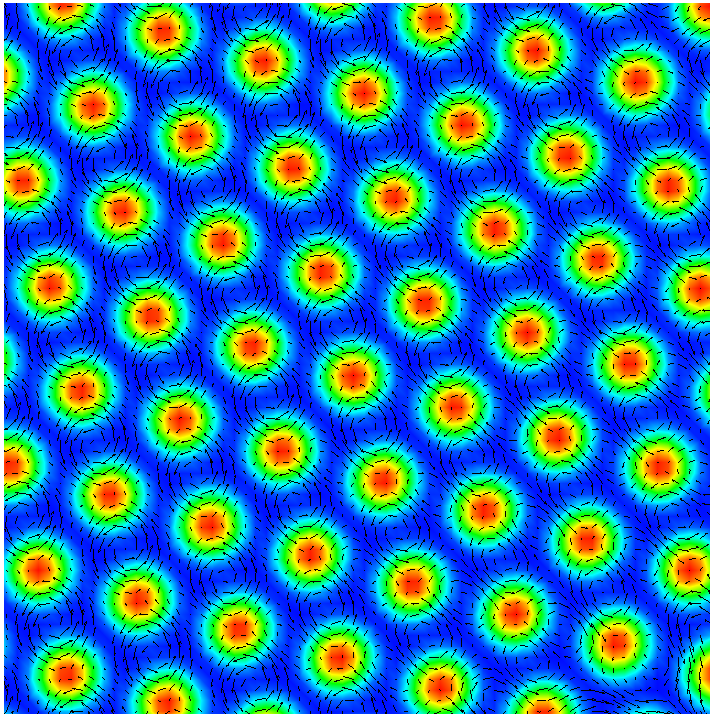}} \
\subfloat[][\emph{$\zeta=0.006$}\label{img:zoom0.006}]
{\includegraphics[width=.3\textwidth]{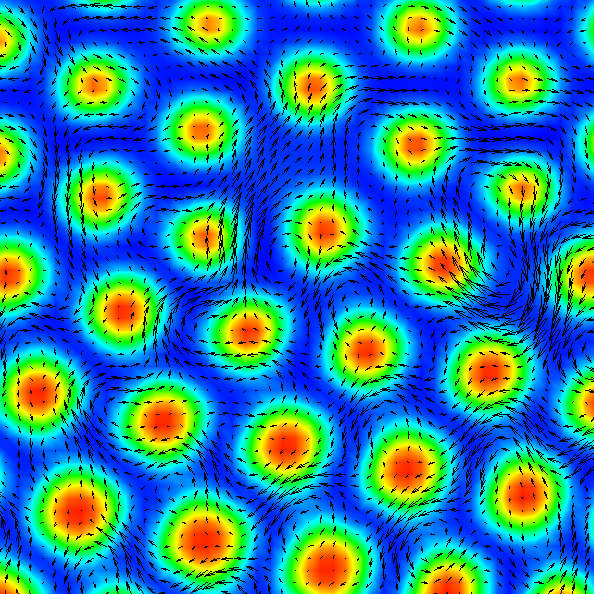}} \
\subfloat[][\emph{$\zeta=0.0072$}\label{img:zoom0.0072}]
{\includegraphics[width=.299\textwidth]{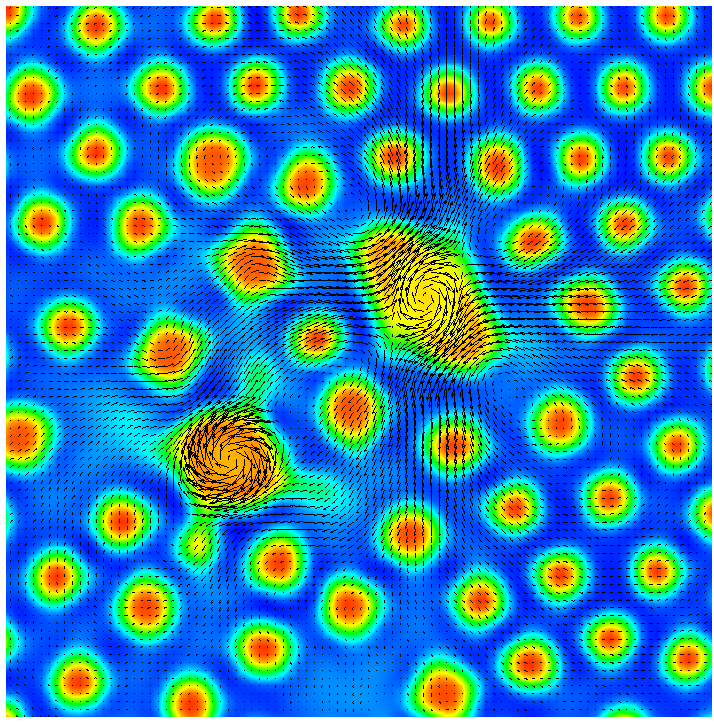}} \
%\subfloat[][\emph{$\zeta=0.05$}\label{img:zoom0.05}]
%{\includegraphics[width=.3\textwidth]{positive_activity/zoom_005.png}}
\caption{Snapshots of $\phi$ contour plots with superimposed the velocity field at late times with different choices of $\zeta>0$, for a portion of size $L=120$ for cases (a) and (c), and of size  $L=100$ for case (b), of a system of original size $L=256$. Averaged modules, over lattice sites, of velocities for the three cases are $3.473\times 10^{-4}$, $7.232\times 10^{-4}$ and $1.974\times 10^{-3}$ respectively.}
\label{fig:contour_plot_extensile_zoom}
\end{figure}

\subsection{Larger activity and aster-like droplets}
%%%%%%%%%%%%%%%%%%%%%%%%%%%%%%%%%%%%%%%%%%%%%%%%%% BEPPE NEW VERSION
Starting from $\zeta=0.0072$ aster-like droplets are no more transient and the system morphology, at late times, becomes dominated by the presence of these structures\footnote{We checked the behavior until $t=3.5\times10^6$ as can be seen from the time evolution of the kinetic energy in Fig.\ref{fig:en.evolution}. }. The transition to this regime can be analyzed, qualitatively, looking at the panels in Fig.~\ref{fig:SEQUENCE_FINO}.
Here  a series of snapshots of $\phi$ contour plots for different values of activity are reported, starting from $\zeta=0.0072$ up to $\zeta=0.0090$.
A small number of isolated aster-like droplets linger at late times for $\zeta=0.0072$ (Fig.~\ref{zeta=0.0072}), while for bigger values of activity these structures start grouping (Fig. \ref{zeta=0.0074}) filling progressively a large portion of the system (Figs.~\ref{zeta=0.0078},~\ref{zeta=0.0080} and~\ref{zeta=0.0090}) at the expense of small droplets.
\begin{figure}[t!]
\centering
{\includegraphics[width=.79\textwidth]{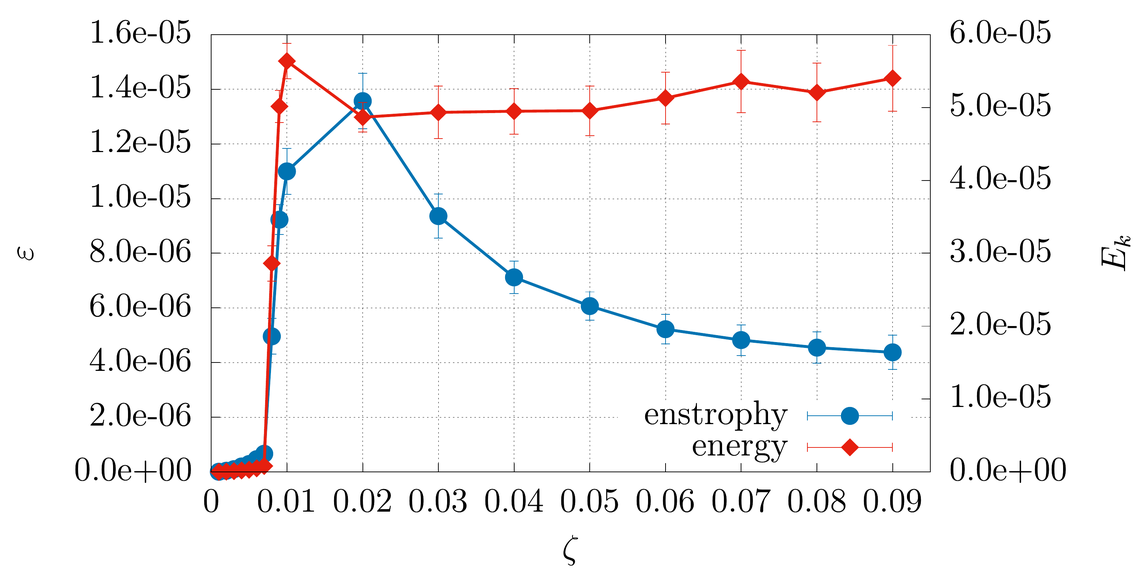}} \
%{\includegraphics[width=.49\textwidth]{positive_activity/enstrophy:vs:zetapos.png}}
\caption{Plot of kinetic energy and enstrophy per unit volume averaged in time as functions of $\zeta>0$.} 
\label{fig:observables_extensile}
\end{figure}
The mechanism of formation of the aster-like droplets can be intuited looking at  Figs.     
 \ref{zeta=0.0072} and \ref{zeta=0.0074}. First, bigger droplets thicken in a portion of the system, leading two of them to merge, then the new big droplet starts rotating while incorporating neighboring droplets until an aster-like structure is fully stable. 
The time evolution of the kinetic energy per unit volume $E_k=\sum_{\mathbf{r}}\rho       
\mathbf{v} (\mathbf{r})^2/2$, for $\zeta=0.0072$, is
shown in Fig.~\ref{fig:en.evolution} (blue curve) and reproduces the aster formation: The system sits in the droplet configuration with a constant value of kinetic energy density until the merging of the two droplets leads to a strong discontinuity in the energy density. 
When the aster-like droplets are fully stable, there is a strong contribution to the kinetic energy coming from their angular velocity. These structures then begin to slow down until they disappear,  and the energy density returns to a lower constant value. Then, again, new asters are formed and do not disappear even at late times. In contrast, at lower activity (yellow curve in Fig. \ref{fig:en.evolution}, for $\zeta=0.006$) asters disappear and the kinetic energy goes to zero  remaining constant for the time checked.
The red curve in Fig. \ref{fig:en.evolution} corresponds to the case $\zeta=0.0078$. The kinetic energy is appreciably greater then the previous cases and increases until a stationary state is reached.  For this value of activity asters are continuously formed and never disappear in the system.

The velocity field at different activities is displayed in Fig.~\ref{fig:contour_plot_extensile_zoom}. We see that  the resulting  aster-like droplets            
(Fig. \ref{img:zoom0.0072}) have a velocity field much stronger that the one around droplets
 (see caption of Fig. \ref{img:zoom0.0072}). They behave as a vortical source for the velocity field and for this reason we refer to them also as big rotating droplets.
The velocity field as activity varies reproduces the behaviors so far discussed.
At  $\zeta=0.006$ (Fig.~\ref{img:zoom0.006}) we distinguish vortical structures that are not localized anymore in correspondence of the droplets, as for smaller values of the activity (Fig. \ref{img:zoom0.0030}), but start mixing with neighboring vortices.  
%%%%%%%%%%%%%%%%%%%%%%%%%%%%%%%%%%%%%%%%%%%%%%%%%%
 \begin{figure}[t!]
\centering
%\subfloat[][ \label{img:Zeta0.09}]
%{\includegraphics[width=.308\textheight]{positive_activity/256_09_bis.png}} \
%\subfloat[][\label{img:vec_Zeta0.09}]
{\includegraphics[width=.408\textheight]{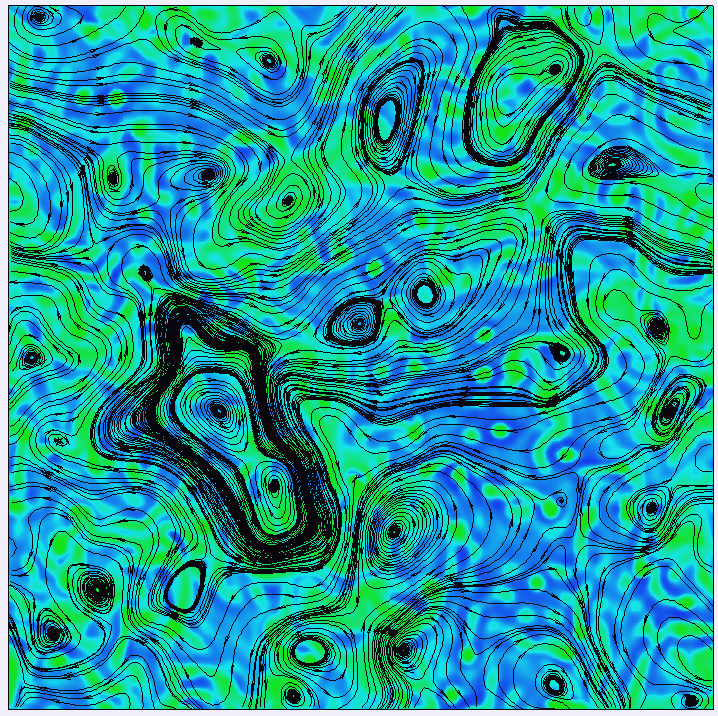}} \\
\caption{Snapshots of $\phi$ contour plot with superimposed streamtraces,  at late times, for $\zeta=0.09$, for a lattice of size $L=256$.The color code is the same of that displayed in Fig. \ref{img:no_activity_phi}. The prevalence of green regions means that the system is almost completely mixed.}
\label{fig:contour_plot_extensile}
\end{figure}

\subsection{Very large activity and overview}
At even larger values of activity ($\zeta\gtrsim0.01$) the aster-like structures are destroyed by the flow and no clear morphological pattern can be observed. The system appears completely mixed as it will be shown.  

A better characterization of the behavior of the system in this regime comes from the study of  hydrodynamic quantities such as the kinetic energy per unit volume 
$E_k$ and the enstrophy per unit volume defined as $\varepsilon=\sum_{\mathbf{r}} \mathbf{\omega}(\mathbf{r})^2/2$, 
where $\omega_i = \epsilon_{ijk} \partial_j \mathrm{v}_k$ is the vorticity vector and $\epsilon_{ijk}$ is the completely anti-symmetric Levi-Civita tensor. Kinetic energy represents a measure of the strength of fluid flows in the system, while enstrophy can be used to figure out whether 
the velocity field has developed vortical structures.
\begin{figure}[t!]
\centering
\subfloat[][\emph{$\zeta=-0.008$} \label{img:-0.003}]
{\includegraphics[width=.32\textwidth]{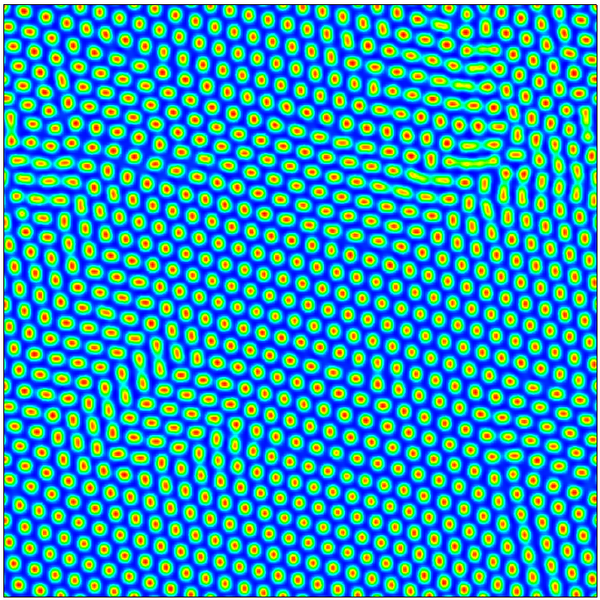}} \quad
%\subfloat[][\emph{$\zeta=-0.01$} \label{img:-0.008}]
%{\includegraphics[width=.3\textwidth]{negative_activity/256_01.png}} \
\subfloat[][\emph{$\zeta=-0.02$} \label{img:-0.013}]
{\includegraphics[width=.32\textwidth]{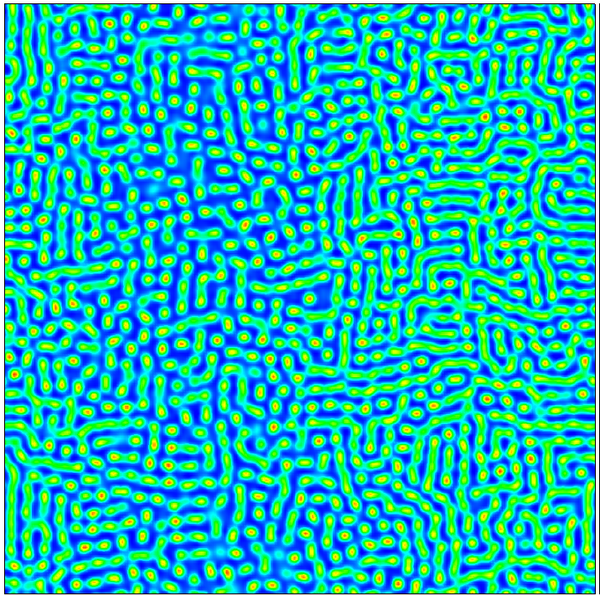}} \\
\subfloat[][\emph{$\zeta=-0.008$}. \label{img:zoom-0.003}]
{\includegraphics[width=.32\textwidth]{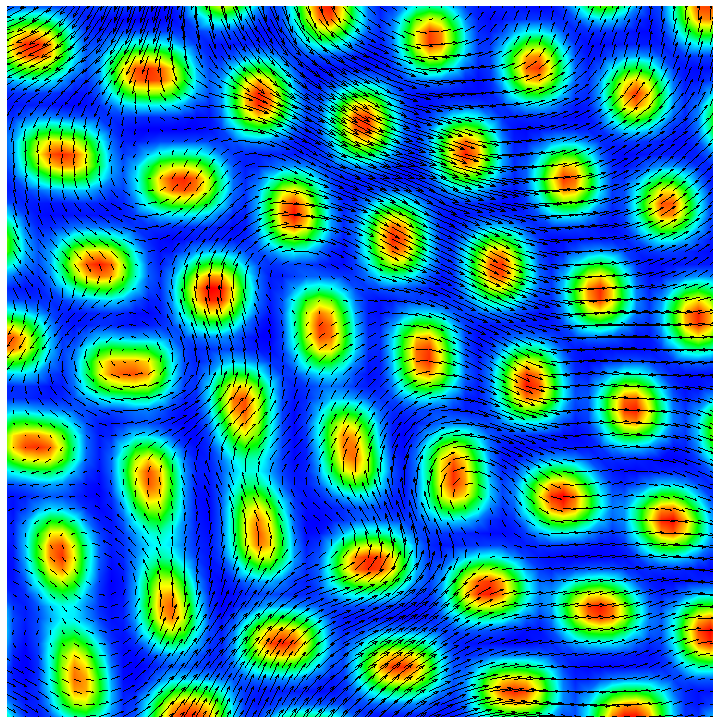}} \quad
\subfloat[][\emph{$\zeta=-0.02$}. \label{img:zoom-0.02}]
{\includegraphics[width=.32\textwidth]{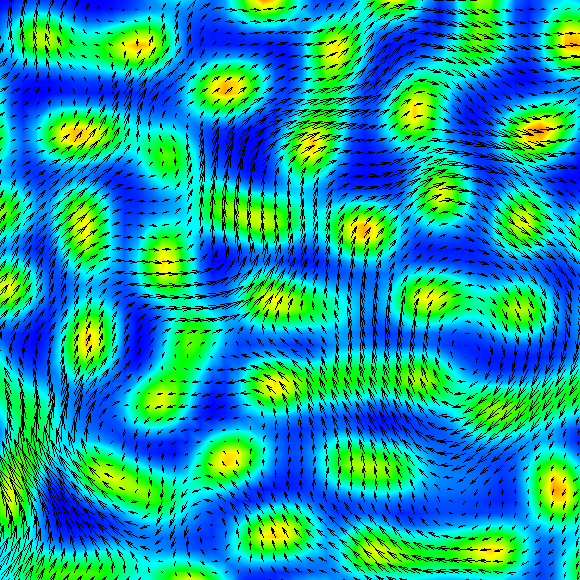}}
\caption{(\textbf{a})-(\textbf{b}) Snapshots of $\phi$ contour plots at late time for different choices of $\zeta<0$, for a system of size $L=256$.
(\textbf{c})-(\textbf{d}) Snapshots of $\phi$ contour plots with superimposed the velocity field at late times, for the same  choices of $\zeta<0$ as in (a) and (b) respectively, for a portion of size $L=70$ of a system of original size $L=256$. Averaged modules, over lattice sites, of velocities for the two cases are $9.201\times 10^{-4}$ and $1.143\times 10^{-3}$ respectively.}
\label{fig:contour_plot_contractile}
\end{figure}
The graphs in Fig.~\ref{fig:observables_extensile} show the behavior of energy and enstrophy averaged in time over uncorrelated configurations as functions of the activity. They are both null for $\zeta=0$ and raise to reach a peak in the range $0.01-0.02$. Values are 
significantly different from $0$ at $\zeta=0.0072$ and this corresponds to the formation of vortical asters.
%The number of vortical structures increases until $\zeta\simeq 0.02$, when
If activity is raised over $\zeta \simeq 0.007$, the number of vortical structures increases until $\zeta\simeq 0.02$. For larger values of $\zeta$ asters become unstable and flexible, elongated rotating structures are formed until one arrives at configurations like that of Fig.~\ref{fig:contour_plot_extensile}, at $\zeta=0.09$, with the velocity field exhibiting a pattern like the one shown.
%and no definite pattern can be finally observed in the system (Fig. \ref{img:Zeta0.09}). 
%we observe the formation of many vortical structures until  
When these structures start melting, rotational contribution to the kinetic energy starts decreasing while the kinetic energy stays approximately constant.
In this regime weak vortical flows span the system. Such flows dissipate energy while moving, according to the dissipative nature of the fluid, and weaken in intensity until they expire in small and slow vortices or simply merge in more intense flows as suggested by the streamtraces displayed in Fig. \ref{fig:contour_plot_extensile}.%, for $\zeta=0.09$. 
\begin{figure}[t!]
\centering
%\subfloat[][ \label{zeta.ext_ener}]
{\includegraphics[width=.8\textwidth]{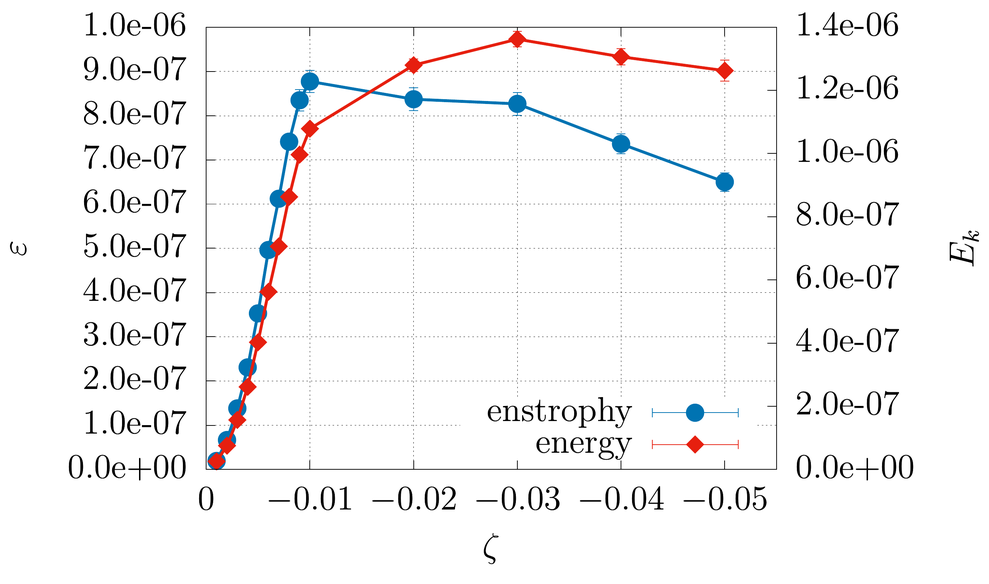}} \
%\subfloat[][ \label{zeta.ext_enstr}]
%{\includegraphics[width=.49\textwidth]{negative_activity/enstrophy:vs:zetaneg.png}} 
\caption{Plot of energy and enstrophy per unit volume as functions of $\zeta$. Both energy and enstrophy are two order of magnitude less compared to the extensile case (Fig. \ref{fig:en.evolution}). This is mainly due to the absence of asters in this case.}
\label{fig:observables_contractile}
\end{figure}
%Surprisingly, for very strong values of activity, the mixing may lead to highly ordered structures, such the one shown in Fig.~\ref{fig:vel_contour_plot_extensile} characterized by vortex in the velocity field centered around them.
%high angular velocity (Fig.~\ref{vel_zeta=0.05}). 

\section{Contractile case}

We now consider the case of contractile ($\zeta<0$) emulsions.
For slightly negative values of $\zeta$, configurations at late times are similar to the extensile case.  
%the hexatic order is preserved 
%in a way that defects in the hexagonal layout diminish 
%with respect to the case with no activity.
%due to the presence of a residual fluid flow sustained by active stress.

Increasing $\abs{\zeta}$, the hexatic order is progressively lost: First a change in the droplet dimension and shape is observed, as it can be seen in Fig.~\ref{img:-0.003}. This leads the fluid flow to assume different configurations throughout the system. At very low contractile activity, not shown, the velocity field is small in magnitude and enforced in regions close to droplets creating vortical structures that rotate around the center of the droplets themselves.
%while it tends to reduce its intensity in the passive (blue) phase. For this reason, in the case $\zeta=-0.003$, regions with bigger droplets are characterized by more intense flow than the ones with smaller droplets.
For values of activity, such as $\zeta=-0.008$, droplets start oscillating, following the flow (Fig. {\ref{img:zoom-0.003}}) and, eventually, merging (see Figs~.\ref{img:-0.003} and \ref{img:zoom-0.003}) thus creating elongated structures.
For large enough  activity  the hexagonal pattern is lost (Fig. \ref{img:-0.013}).
Flow fields are sustained by the active stress that continuously drives energy into the system and generates flows that are not confined around the droplets anymore, but move throughout 
the system as it is shown in Fig.~\ref{img:zoom-0.02}.
This happens for values of $\zeta$ ranging from $-0.01$ to $-0.02$ but for even stronger values, the merging of structures throughout the system affects the equilibrium configuration so strongly that when $\zeta  \lesssim   -0.02$, it is not possible to distinguish any kind of definite pattern anymore.  

%\begin{figure}[t!]
%\centering
%{\includegraphics[width=.38\textheight]{positive_activity/vfb_05.png}}\\
%\caption{Snapshots of $\phi$ contour plots and velocity field  for a portion of size $L=100$  at $\zeta=0.05$. }
%\label{fig:vel_contour_plot_extensile}
%\end{figure}

Here again, together with the morphology it is useful to look at the behavior of the kinetic energy and enstrophy per unit volume. The graphs in Fig.~\ref{fig:observables_contractile} show their behaviors averaged in time over uncorrelated configurations of the system for different values of the activity $\zeta$. Both energy and enstrophy are null when $\zeta=0$, according to the fact that in absence of an energy source no fluid flow can be sustained in a dissipative fluid; then they increase until $\zeta \simeq -0.01$ when
enstrophy reaches a peak. This trend fits the morphology behavior presented before: As far as droplets are preserved, the velocity field creates vortical structures around them that raise in strength for stronger values of activity, but when the merging of droplets starts affecting the morphology, the local vortical structures are progressively replaced by flows without ordered structure. 
This brings energy to stay approximatively
constant for values $\zeta \lesssim -0.02$ while enstrophy collapses rapidly.

%It is worth noticing that the energy and enstrophy in the contractile case are two order of magnitude less compared to the extensile case. This is mainly due to the fact that in the former case asters-like droplets, with high angular velocity, do not compare.  

%\begin{figure}
%\centering
%\subfloat[][\emph{$\zeta=-0.003$}. \label{img:zoom-0.003}]
%{\includegraphics[width=.4\textwidth]{negative_activity/zoom_-0002.png}} \quad
%\subfloat[][\emph{$\zeta=-0.02$}. \label{img:zoom-0.02}]
%{\includegraphics[width=.4\textwidth]{negative_activity/zoom_-002.png}}
%\caption{Snapshots of $\phi$ contour plots with superimposed the velocity field at late times with different choices of $\zeta<0$, for a portion of a system of original size $L=256$. }
%\label{fig:contour_plot_contractile_zoom}
%\end{figure}

\section{Conclusions and perspectives}
%We have analyzed the behavior of a highly asymmetric exotic active emulsion for different values of the activity both in the extensile and in the contractile case.
%In the absence of activity, at equilibrium the system is in a hexatic ordered phase with some defects in the arrangement.
%Small amount of activity, both in the contractile and in the extensile case, preserves the droplet ordered phase. 
%For strong active doping the hexatic order is progressively lost with some differences between the two cases. 
%We have characterized the dynamics of this process looking at the energy and the enstrophy evolution. 
%\begin{figure}[t!]
%\centering
%\subfloat[][\emph{$\zeta=-0.008$}. \label{img:zoom-0.003}]
%{\includegraphics[width=.4\textwidth]{negative_activity/zoom_-0008_bis.png}} \quad
%\subfloat[][\emph{$\zeta=-0.02$}. \label{img:zoom-0.02}]
%{\includegraphics[width=.4\textwidth]{negative_activity/zoom_-002.png}}
%\caption{Snapshots of $\phi$ contour plots with superimposed the velocity field at late times with different choices of $\zeta<0$, for a portion of size $L=70$ of a system of original size $L=256$. Averaged modules, over lattice sites, of velocities for the two cases are $9.201302\times 10^{-4}$ and $1.143674\times 10^{-3}$ respectively.}
%\label{fig:contour_plot_contractile_zoom}
%\end{figure}
This work shows how activity modifies the equilibrium hexatic-ordered droplets configuration of an highly asymmetric active emulsion. 

In the extensile case we found three different regimes. For $\zeta$ up to $0.005$ hexatic order is preserved, while is enhanced for values of $\zeta$ around $0.006$, which is the critical value for a transition to a regime dominated by  big-rotating droplets. In fact, starting from $\zeta=0.0072$ the system morphology becomes characterized by the presence of aster-like droplets. At even larger values of activity ($\zeta\gtrsim0.01$) the aster-like structures are destroyed by the flow and the system appears completely mixed.
We have scored the transitions studying energy and enstrophy behaviors.
% For larger values of activity the %velocity field forms vortices no longer centered on the droplets giving rise to the formation of 
%isolated structures and destroying any order throughout the mixture. 
In the case of contractile emulsions, for slightly negative values of $\zeta$ stationary configurations are similar to the extensile case while, strong contractile activity gives rise to elongated structures.

In the future we plan to extend these studies to three dimensional systems, where even richer morphologies ad flow patterns ca be expected.

%The study here presented is the necessary continuation of the work of Bonelli \emph{et al.}~\cite{bone2017}. They showed that a variety of morphologies can be obtained even for a symmetric mixture ranging from self-assembling of a suspension of active rotating droplets to enhancement of lamellar order and formation of an emulsion of passive droplets in an active background, respectively in extensile and contractile systems.
%The unexpected features exhibited by this new kind of materials and the possibility of tuning the properties of the system by varying the intensity of active doping, may be useful in designing new microfluid devices whose behavior can be kept under control according to the amount of fuel available to the active component.
%This would also represent a further important step in the study of active turbulent fluids. %\cite{thampi2016active,PhysRevX.5.031003} (with the possibility of tuning the intensity but also the spatial distribution of energy input in the system).

\section{Acknowledgments}
Simulations have been performed at Bari ReCaS e-Infrastructure
funded by MIUR through the program PON Research and
Competitiveness 2007-2013 Call 254 Action I. We thank D. Marenduzzo, A. Tiribocchi and F. Bonelli for the very useful discussions.

%%%%%%%%%%%%%%%%%%%%%%%%%%%%%%%%%%%%%%%%%%%%%%%%%%%%%%%%
\clearpage

%\begin{appendices}
\appendix
\section{Numerical method}{\label{appendix1}}
In the LB scheme the evolution of the fluid is defined in terms of a set of $N$ discrete distribution functions $\{f_i\}$ $(i=0,...,N-1)$ which obey the dimensionless Boltzmann equation in the BGK approximation:
\begin{equation}
\begin{split}  
 & f_i({\bf{r}} +{\bf{e}}_i \Delta t , t+\Delta t)-f_i({\bf{r}}, t)\\
 &=-\frac{\Delta t}{\tau}[ f_i({\bf{r}}, t)-f_i^{\textit{eq}}({\bf{r}}, t)] \ ,
\end{split}
\end{equation}
where $\mathbf{r}$ and $t$ are the spatial coordinates and the time, respectively, $\{\bf{e_i}\}$ $(i=0,...,N-1)$ is the set of discrete velocities, $\Delta t$ is the time step, and $\tau$ is a relaxation time which characterizes the relaxation towards the equilibrium distributions $f_i^{\textit{eq}}$. The shear viscosity $\eta$ is related to $\tau$ by the relationship $\eta=c^2 \Delta t \frac{2\tau - 1}{6}$. The value of $N$ depends on the space dimensions
and the lattice geometry.
The moments of the distribution functions allow to write the conservation laws for the density and total momentum in the form:
\begin{equation}\label{m1}
\begin{split}
&\sum_i f_i^{\textit{eq}}=\rho\\
&\sum_i f_i^{\textit{eq}}e_{i\alpha}=\rho \mathbf{v}_{\alpha} \ .
\end{split}
\end{equation}

The second moment, which describes the balance between energetic densities at stake, is fixed in order to find the hydrodynamic equations in the continuum limit. It is given by 

\begin{equation}\label{m2}
\sum_i f_i^{\textit{eq}} e_{i\alpha} e_{i\beta}=\Pi_{\alpha\beta}+\rho \mathbf{v}_{\alpha} \mathbf{v}_{\beta} \ ,
\end{equation}
where $\Pi_{\alpha\beta}$ represents the pressure tensor. Introducing the nematic tensor $Q_{\alpha\beta}={\bf P}_{\alpha}{\bf P}_{\beta}-\frac{1}{3} |{\bf P}^2|\delta_{\alpha\beta}$, the pressure tensor can be written as
\begin{equation}
\begin{split}
\Pi_{\alpha\beta}&=-p\delta_{\alpha\beta}+2\xi(Q_{\alpha\beta}+\frac{1}{3}\delta_{\alpha\beta})Q_{\gamma\epsilon}H_{\gamma\epsilon}\\
&-\xi H_{\alpha\gamma}(Q_{\gamma\beta}+\frac{1}{3}\delta_{\gamma\beta})-\xi(Q_{\alpha\gamma}+\frac{1}{3}\delta_{\alpha\gamma})H_{\gamma\beta}\\
&-\partial_{\alpha}Q_{\gamma\nu}\frac{\delta f}{\delta(\partial_{\beta}Q_{\gamma\nu})}+Q_{\alpha\gamma}H_{\gamma\beta}-H_{\alpha\gamma}Q_{\gamma\beta}\ ,
\end{split}
\end{equation}
where $H$ is the field conjugated to the nematic tensor and $p$ the ideal gas pressure.

The equilibrium distributions are expanded up to the second order in the velocities
\begin{equation}
\begin{split}
  f_k^{\textit{eq}}=& A_k+B_k \mathbf{v}_\alpha e_{k\alpha} + C_k\mathbf{v}^2+\\
  & D_k\mathbf{v}_{\alpha}\mathbf{v}_{\beta}e_{k\alpha}e_{k\beta}+G_{k\alpha\beta}e_{k\alpha}e_{k\beta}\ ,
\end{split}
\end{equation}
where the index $k$ labels the different directions on the discretized  lattice.
%\textcolor{black}corresponding to the different modules of the lattice velocity $0,u,\sqr
The $A_k$, $B_k$, $C_k$, $D_k$ and $G_{k\alpha \beta}$ are characteristic parameters to be determined to get the right hydrodynamic equations in the continuum limit by imposing the conditions given by Eqs.~(\ref{m1})-(\ref{m2}).
For a two-dimensional square lattice with $N=9$ velocities (D$2$Q$9$), which is the model here considered, these parameters are reported in Table \ref{cof}, and the lattice velocities are ${\bf e}_0=(0,0)$, ${\bf e}_{1,2}=(\pm u,0)$, ${\bf e}_{3,4}=(0,\pm u)$, ${\bf e}_{5,6}=(\pm u,\pm u)$, ${\bf e}_{7,8}=(\mp u,\pm u)$ with $u=\Delta x/\Delta t$, $\Delta x$ being the lattice mesh size.
 In Table \ref{cof} $k=0$ corresponds to the rest lattice velocity, while $k=1$ and  $k=2$ correspond to the velocities directed towards the first and second neighbors, respectively.

%%%%%%%%%%%%%%%%%Table coefficients
\begin{table}
\caption{Equilibrium distribution coefficients}\label{cof}
\centering
\begin{tabular}{ |l | c | r|}
\hline\hline
 $A_0=\rho-20 A_2$ &   $A_1=4A_2$ & $A_2=\frac{[\Pi_{\alpha\beta}+\mathbf{v}\zeta(Q_{\alpha\beta}+\frac{1}{3}\delta_{\alpha\beta})]\delta_{\alpha\beta}}{24\mathbf{v}^2}$\\
 \hline 
$B_0=0$ & $B_1=4B_2$  &   $B_2=\rho/12 \mathbf{v}^2$\\
\hline  
$C_0=-2\rho/3\mathbf{v}^2$  &   $C_1=4C_2$ & $C_2=-\rho/24 \mathbf{v}^2$ \\
\hline 
$D_0=0$ & $D_1=8 D_2$ &   $D_2=\rho/8\mathbf{v}^2$\\
\hline 
$G_{0\alpha\beta}=0$ & $G_{1\alpha\beta}=4 G_{2\alpha\beta}$  & $G_{2\alpha\beta}=\frac{[\Pi_{\alpha\beta}-\frac{1}{2}\Pi_{\delta\delta}\delta_{\alpha\beta}+\mathbf{v}\zeta(Q_{\alpha\beta}+\frac{1}{3}\delta_{\alpha\beta})]}{8\mathbf{v}^4}$\\
\hline\hline
\end{tabular}
\end{table}

The LB scheme is coupled with a finite-difference predictor-corrector algorithm, in order to solve simultaneously Eqs.~(\ref{nav})-(\ref{P_eq}).
This method was previously used to study the hydrodynamics of binary fluids~\cite{tiri2009,hybrid_2}, liquid crystals and active matter \cite{cate2009,Bonelli}. 

%\end{appendices}
%%%%%%%%%%%%%%%%%%%%%%%%%%%%%%%%%%%%%%%%%%%%%%%%%%%%%%%%%%%%%FINE APPENDICI

\clearpage

\bibliographystyle{unsrt}

\bibliography{refs}

\end{document}